%% file: main.tex
\documentclass[sigplan,nonacm]{acmart}

\usepackage[]{hyperref}
\usepackage{tikz}

\usepackage{algorithmic}
\usepackage{graphicx}
\usepackage{textcomp}
\usepackage{xcolor}
\usepackage{amsmath}
\usepackage{pgfplots}
\usepackage{geometry}
\usepackage{array}
\usepackage{booktabs}
\pgfplotsset{compat=1.18}
\usetikzlibrary{shapes.geometric, arrows.meta, positioning}
\usetikzlibrary{shapes, arrows}
\usetikzlibrary{positioning, calc} 
\usetikzlibrary{decorations.pathmorphing} %
\usepackage{multirow}
\usepackage{subfig}
\usepackage{subcaption}
\usepackage{listings}

\newcommand{\mytitle}[1]{\textcolor{black}{GPUVM}}

\begin{document}
\title{GPUVM: \underline{G}\underline{P}U-driven \underline{U}nified \underline{V}irtual \underline{M}emory}

\author{
    \begin{tabular}{ccc}
        {\rm Nurlan Nazaraliyev} & & \hspace{-1.5cm} {\rm Elaheh Sadredini} \\
        \textit{University of California Riverside} & & \hspace{-1.5cm} \textit{University of California Riverside} \\
        & \hspace{-1.5cm} {\rm Nael Abu-Ghazaleh} & \\
        &  \hspace{-1.5cm} \textit{University of California Riverside} &
    \end{tabular}
}

% ...
\input{Abstract}

\maketitle 
\thispagestyle{firstpage}
\pagestyle{plain} 


\vspace{-0.2in}
\section{Introduction}
\input{introduction}

\section{Background}
\input{background}

\section{\mytitle{} System Design and Architecture}
\input{system}

\input{table_hardware}

\vspace{-0.2in}
\section{\mytitle{} Implementation}
\input{prototype}

\section{Evaluation}
\input{evaluation}

\subsection{Discussion}
\input{discussion}
\section{Related Work}
\input{related_work}
\section{Concluding Remarks}
\input{conclusion}

\bibliographystyle{plain}
\input{ref}

\end{document}

%% file: Abstract.tex
\begin{abstract}
  Graphics Processing Units (GPUs) leverage massive parallelism and large memory bandwidth to support high-perform\-ance computing applications, such as multimedia rendering, crypto-mining, deep learning, and natural language processing.   These applications require models and datasets that are getting bigger in size and currently challenge the memory capacity of a single GPU, causing substantial performance overheads. 
  To address this problem, a programmer has to partition the data and manually transfer data in and out of the GPU.  This approach requires programmers to carefully tune their applications and can be impractical for workloads with irregular access patterns, such as deep learning, recommender systems, and graph applications. To ease programmability, programming abstractions such as unified virtual memory (UVM) can be used, creating a virtually unified memory space across the whole system and transparently moving the data on demand as it is accessed.  However, UVM brings in the overhead of the OS involvement and inefficiencies due to generating many transfer requests especially when the GPU memory is oversubscribed.  
  This paper proposes \mytitle{}, a GPU memory management system that uses an RDMA-capable network device to construct a
  virtual memory system without involving the CPU/OS. 
  \mytitle{} enables on-demand paging for GPU applications and relies on GPU threads for memory management and page migration. Since CPU chipsets do not support GPU-driven memory management, we use a network interface card to facilitate transparent page migration from/to the GPU. 
\mytitle{} achieves performance up to 4$\times$ higher than UVM for latency-bound applications while providing accessible programming abstractions that do not require the users to manage memory transfers directly. 

\end{abstract}

%% file: introduction.tex

\sloppy
GPUs are popular platforms for accelerating data-intensive high-performance applications, such as graph analytics~\cite{jiang2024core, afarin2023commongraph, chen2023compressgraph, wang2021grus}, recommender systems~\cite{djenouri2023efficient}, machine learning~\cite{hu2021characterization} and natural language processing~\cite{sheng2023flexgen, jiang2024megascale}. GPUs use massively parallel, high-throughput architectures that can provide high computing performance and memory bandwidth~\cite{A100, choquette2023nvidia}. 
GPUs are widely employed to accelerate data-intensive applications, which frequently operate on large-scale datasets, typically ranging in size from several gigabytes to tens of terabytes, and are likely to continue to increase in size in the foreseeable future. 
The memory demands of modern data-intensive applications continue to outpace available GPU memory, despite increases in GPU memory capacity (e.g., up to 80 GBs for NVIDIA A100 GPUs ~\cite{nvidia_dgx_a100_2020} and 94GB for NVIDIA H100 GPUs ~\cite{nvidia_h100_2024}).  In such cases, the application's memory does not fit on the GPU and typically resides on the CPU and
 programmers are responsible for moving data back and forth to the GPU (using \textit{cudaMemCpy} or similar APIs), to ensure that the required data are available at the GPU when needed.   This approach substantially complicates programming and requires careful optimization of data transfer operations, which is often not possible when applications are irregular with difficult-to-predict memory access patterns.

To ease these programming burdens, NVIDIA introduced the Uniform Virtual Memory (UVM) ~\cite{UVM} abstraction, where memory migration is automatically managed during run time. While UVM improves programmability and portability, it leads to substantial overheads~\cite{allen2021demystifying, allen2021depth, wagley2024exploring}. 
Specifically, the performance of UVM and prior related works ~\cite{kim2020batch, yu2020coordinated, go2023early} is limited for generating as many requests as to utilize the available PCIe bandwidth fully.
The latencies associated with handling transfer requests, as well as the inefficient use of available bandwidth which is an order of magnitude or lower than the GPU memory bandwidth ~\cite{nvidia_dgx_a100_2020} can substantially limit performance when the working memory does not fit within the GPU memory.  We discuss the shortcomings of current approaches for extending the effective size of the GPU memory in more detail in Section~\ref{sec:background}. 
As a result, a programmer is left with an unsatisfactory choice of (1) Sacrificing programmability and manually managing and optimizing memory transfer operations between the CPU and GPU.  This approach is cumbersome, hardware-specific, and may not be practical for applications where the memory access patterns are irregular; or (2) Sacrificing performance and using UVM to ease programmability and portability by having the system manage the data transfers.   In this paper, we introduce GPUVM, a new GPU memory management system that supports both programmability and performance.  
It allows GPUs to directly manage memory, using Remote Direct Memory Access (RDMA) ~\cite{ambal2024semantics} to substantially reduce latency and improve data transfer throughput. GPUVM eliminates CPU involvement and the latencies associated with CPU-mediated page fault handling.  It also enables parallel fault handling to allow operations to overlap with ongoing GPU computations, thus improving overall system efficiency. 
GPUVM incorporates several optimizations, including coalescing related data transfers and overlapping them with computation. 
In addition, GPUVM employs highly efficient memory management and eviction schemes that deliver even higher advantages with increasing pressure on the GPU memory.

GPUVM has to solve a number of challenges arising from the model and the limitations of current hardware. On the GPU side, we have to build a runtime library that handles page faults, manages the available memory space, and initiates communication to transfer memory pages. Specifically, GPUVM implements a high-throughput, highly parallel, synchronous, and low-latency on-demand paged memory system to address this issue. The paged-memory system is optimized to coalesce the access requests and reuse the fetched pages.  To handle oversubscribed cases where the application memory exceeds the available GPU memory, GPUVM implements eviction logic to create room for newly accessed pages. We note that current CPU chipsets do not support GPU-managed data transfers between CPU and GPU memory. To overcome this challenge, GPUVM uses a network interface card to facilitate establishing RDMA connections which are used to mediate the transfer of the data from the CPU to the GPU. We provide more information about the design and implementation of GPUVM in Section~\ref{sec:system_overview}.

We implement the GPUVM software stack\footnote{We will go through the artifact evaluation and release the source code.} and evaluate it using Cloudlab ~\cite{Duplyakin+:ATC19}.  We demonstrate GPUVM on a number of benchmarks, demonstrating how it can directly support a range of applications without substantial programming overhead.  GPUVM achieves substantially higher performance than UVM on a number of GPU benchmarks with multiple datasets. It also outperforms optimized graph frameworks such as Subway~\cite{sabet2020subway}.



In summary, the contributions are as follows.
\begin{itemize}

    \sloppy
    \item We present GPUVM, a new model for managing GPU virtual memory, supporting demand paging without CPU intervention. GPUVM leverages GPUDirect RDMA to enable GPUs to manage memory management directly, substantially lowering the latency of page faults and improving PCIe utilization.
    
    
    \item We develop a high-level programming abstraction and software APIs to facilitate the integration of GPUVM into existing applications.

    \item To support larger applications, we build a reuse-oriented paged memory allowing efficient eviction of pages when there is memory pressure on GPUs.
      
    \item We evaluate GPUVM on a number of benchmarks, showing that it substantially outperforms existing solutions including NVIDIA's UVM. GPUVM achieves up to 4$\times$ 
    speedup, which increases with the degree of memory pressure on the GPU.  

    \end{itemize}

%% file: background.tex
\label{sec:background}
In this section, we provide background on currently used paged memory systems for GPUs; these are the systems that automate the management of memory across GPUs and the host CPU. We then explore how employing RDMA to manage virtual memory can alleviate the GPU memory wall problem and hide the memory access latency by allowing computations to be coalesced with memory operations.

\input{uvm_diagram}

\subsection{Unified Virtual Memory (UVM)}
\label{sec:existing_approaches}
Unified virtual memory (UVM) creates a common virtual address space shared among all available memories in the system. The address space is used when memory is allocated using ~\textit{cudaMallocManaged} ~\cite{UVM}. For a system with a single GPU and CPU (host) shown in Figure ~\ref{fig:uvm_details}, the allocation results in two separate page tables: one in GPU memory and the other in host/CPU memory. The UVM driver manages these page tables and handles page faults, both those originating from the CPU side, or from the GPU side through the PCIe bus. The simplified page fault workflow from the GPU side is shown in Figure ~\ref{fig:uvm_details}. When a thread accesses a remote page (a page residing in host memory), the access first checks the $\mu tlb$ ~\cite{nvidia_open_gpu_doc}: this is a hardware unit that caches recent page translations through multi-level TLBs often consisting of
private Simultaneous Multiprocessor (SM-) and SM-group level TLBs, and GPU-wide shared TLBs shown together in Figure ~\ref{fig:uvm_details} for simplicity. When there is a TLB miss the GMMU (GPU Memory Management Unit) is notified (1), which in turn writes the fault information into a fault buffer (2). Each request sends a hardware interrupt to the UVM driver (3), through a PCIe transaction. As the UVM driver retrieves a batch of faults from the fault buffer (3), it caches them in the host memory and initializes pages in the host memory (4). The UVM driver then informs the OS (5) to handle page table updates (on both host and GPU) and TLB shootdown (on the host) (6). The host OS then directs the DMA engine to migrate the pages to GPU memory (7). 
Previous work has shown that the design of the UVM driver introduces delays in GPU application performance~\cite{allen2021demystifying, yu2020quantitative, allen2021depth}. 
These delays are exacerbated by the lack of parallelism in CPU/OS in handling many requests coming from massively parallel GPU threads.
We also analyzed the overheads of host involvement in page fault handling to be up to 7$\times$ that of page transfer time, as shown in Figure ~\ref{fig:latency_bar}, even at fairly large transfer sizes. 

\input{latency_bar}

\subsection{Heterogeneous Memory Management (HMM)} 

HMM ~\cite{larabel2021amd, sakharnykh2019memory, nvidia_hmm_2020} is a Linux kernel feature designed to simplify memory management between different processing units, such as CPUs and accelerators. HMM provides programmer-agnostic memory management and on-demand memory access by allowing transparent page migration. Unlike UVM, HMM needs no vendor-specific driver and is hardware-agnostic. However, HMM still involves the OS for page fault handling. Both UVM and HMM are based on 4KB pages for $x86\_64$ systems. Conversely, HMM does not support speculative prefetching, an optimization allowed by UVM that asynchronously migrates large GPU page (multiples of 64KB), to hide some of the page fault latency.

\subsection{RDMA support and alternatives} 
\label{sec:background_rdma}



Accelerator devices, such as GPUs, currently lack the capability to initiate data transfers from host memory to device memory due to their limited access to the operating system. To enable device-initiated memory access and data transfers to/from device memory, our approach leverages RDMA-capable network interface cards (RNICs\footnote{In this paper, we use RNIC and NIC interchangeably.}). This solution bypasses the need for CPU involvement, allowing the GPU to directly interact with the NIC for efficient data movement between host memory and device memory.

RNICs have limited capability processors ~\cite{zuo2021one} and enable direct access to host memory and other memories on remote machines across the network ~\cite{li2023rolex}. RDMA can access both local to a single system and remote memory over the network (InfiniBand, Ethernet/RoCE, etc.) ~\cite{gu2017efficient}. Using RDMA read and write requests, a one-sided RDMA connection can be set up in which RNICs can read from and write to application memory without OS/kernel involvement. With GPUDirect RDMA ~\cite{GPUDirectRDMA}, RNICs can also access GPU memory, enabling RNICs to mediate the transfer between a CPU and GPU. 





%% file: uvm_diagram.tex
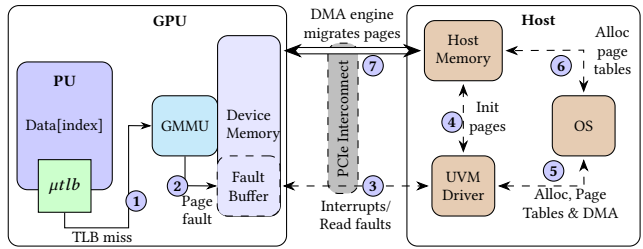
\begin{figure}[t]
    \begin{tikzpicture}

        \tikzset{
            sm/.style={rectangle, rounded corners, draw, text centered, minimum width=3cm, minimum height=1.5cm, fill=blue!20},
            smallrect/.style={rectangle, draw, minimum width=2.5cm, minimum height=1cm, fill=green!20},
            cpu/.style={rectangle, rounded corners, draw, text centered, minimum width=3cm, minimum height=3cm}, 
            smallnode/.style={rectangle, draw, minimum width=1cm, minimum height=1cm, fill=orange!20, text centered}, 
            node2/.style=node2/.style={rectangle, draw, fill=gray!50, minimum width=1.5cm, minimum height=0.8cm, rotate=90, text centered}, 
            ->/.style={draw, -{Stealth[length=1.5mm, width=1mm]}} 
        }

        \node (sm) [sm, minimum height=1.7cm, minimum width=1.2cm] {\tiny Data[index]};

        \node (tlb) [smallrect, below of=sm, yshift=0.2cm, minimum height=0.7cm, minimum width=0.7cm] {\tiny $\mu tlb$};

        \node (gmmu) [sm, right of=sm, xshift=0.6cm, minimum height=0.8cm, minimum width=0.8cm, fill=cyan!20, text=black] {\tiny GMMU};

        \node (devicememory) [sm, right of=gmmu, yshift=0.0cm, xshift=-0.15cm, minimum height=2.4cm, minimum width=0.8cm, fill=blue!10, text=black] {\tiny \parbox[c]{0.6cm}{\centering Device \\ Memory}};

        \node (faultbuffer) [sm, right of=gmmu, xshift=-0.155cm, yshift=-0.80cm, minimum height=0.8cm, minimum width=0.8cm, line width=0.0mm, dashed, fill=blue!10, text=black] {\tiny \parbox[c]{0.6cm}{\centering Fault \\ Buffer}};

        \node (cpu) [cpu, right of=sm, xshift=5.1cm, minimum height=3.2cm, minimum width=3.1cm] {};

        \node (gpu) [cpu, right of=sm, xshift=0.1cm, minimum height=3.2cm, minimum width=3.7cm] {};

        \node (uvm) [sm, xshift=0.2cm, yshift=-0.8cm, minimum height=0.8cm, minimum width=0.8cm, left of=cpu, fill=brown!40, text=black] {\tiny \parbox[c]{0.6cm}{\centering UVM Driver}};

        \node (os) [sm, xshift=1.8cm, yshift=0.0cm, minimum height=0.8cm, minimum width=0.8cm, left of=cpu, fill=brown!40, text=black] {\tiny \parbox[c]{0.6cm}{\centering OS}};

        \node (hostmemory) [sm, xshift=0.2cm, yshift=1.0cm, minimum height=0.8cm, minimum width=0.75cm, left of=cpu, fill=brown!40, text=black] {\tiny \parbox[c]{0.8cm}{\centering Host Memory}};

        \node (pcie) [sm, right of=sm, xshift=2.7cm, yshift=0.1cm, minimum height=0.4cm, minimum width=2.0cm, rotate=90, fill=lightgray, dashed] {\tiny PCIe Interconnect};

        \node at (6.3, 1.4) {\tiny \textbf{Host}}; 
        \node at (1.4, 1.4) {\tiny \textbf{GPU}}; 

        \node at (0.0, 0.6) {\tiny \textbf{PU}};

        \draw[->] (tlb.south) -- ++(0,-0.2) -- ++(0.85,0) -- ++(0,1.36) -- (gmmu.west) 
            node[midway, below=28pt, yshift=-0.3cm, xshift=-0.5cm] {\tiny \parbox[c]{2.0cm}{\centering TLB miss}};
        \node at (0.95, -1.0) [circle, draw, fill=blue!20, inner sep=1pt] {\tiny \textbf{1}};
        
        \draw[->] (gmmu.south) -- ++(0,-0.39) -- (faultbuffer.west)
            node[midway, below, yshift=0.0cm, xshift=-0.05cm] {\tiny \parbox[c]{0.6cm}{\centering Page fault}};
        \node at (1.5, -0.8) [circle, draw, fill=blue!20, inner sep=1pt] {\tiny \textbf{2}};

        \draw[<->, dashed, shorten <=1pt, shorten >=1pt, >=Stealth] (faultbuffer.east) -- (uvm.west)
    node[midway, below=2pt] {\tiny \parbox[c]{1.0cm}{\centering Interrupts/ Read faults}};
        \node at (4.1, -0.8) [circle, draw, fill=blue!20, inner sep=1pt] {\tiny \textbf{3}};


        \draw[<->, double, double distance=2.0pt, shorten <=1pt, shorten >=1pt, >=Stealth] 
          (hostmemory.west) -- ($(devicememory.east) + (0,1.0)$) 
          node[midway, above] {\tiny \parbox[c]{1.5cm}{\centering DMA engine migrates pages}}; 
          \node at (5.15, 0.05) [circle, draw, fill=blue!20, inner sep=1pt] {\tiny \textbf{4}};

        \draw[<->, dashed, shorten <=1pt, shorten >=1pt, >=Stealth] (uvm.east) -- ++(1.18,0) -- (os.south) 
            node[midway, below=3pt, xshift=-5pt] {\tiny \parbox[c]{1.5cm}{\centering Alloc, Page Tables \& DMA}}; 
        \node at (6.5, -0.6) [circle, draw, fill=blue!20, inner sep=1pt] {\tiny \textbf{5}};
            
        \draw[<->, dashed, shorten <=1pt, shorten >=1pt, >=Stealth] (os.north) -- ++(0,0.6) -- (hostmemory.east)
            node[midway, right, xshift=0.4cm] {\tiny \tiny \parbox[c]{0.8cm}{\centering Alloc page tables}};
        \node at (6.6, 0.8) [circle, draw, fill=blue!20, inner sep=1pt] {\tiny \textbf{6}};
        
        \draw[<->, dashed, shorten <=1pt, shorten >=1pt, >=Stealth] (uvm.north) -- (hostmemory.south) 
                node[midway, right, xshift=-0.2cm] {\tiny \tiny \parbox[c]{0.8cm}{\centering Init pages}};
        \node at (4.1, 0.8) [circle, draw, fill=blue!20, inner sep=1pt] {\tiny \textbf{7}};

    \end{tikzpicture}
    \caption{ UVM architecture. ~\textbf{PU} refers to GPU processing units (Streaming Multiprocessors, or \textbf{SMs} on NVIDIA GPUs}
    \label{fig:uvm_details}
\end{figure}

%% file: latency_bar.tex
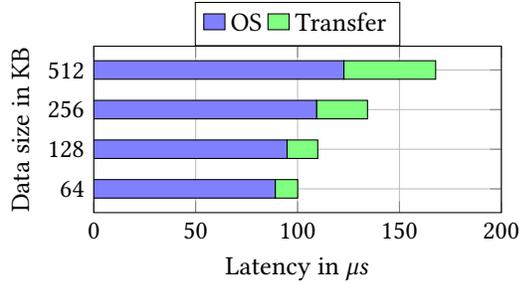
\begin{figure}[t]
    \centering
    \begin{tikzpicture}
        \begin{axis}[
            xbar stacked, 
            bar width=7pt, 
            width=7cm, 
            height=2.0cm, 
            symbolic y coords={64, 128, 256, 512}, 
            ytick=data, 
            xlabel={Latency in $\mu s$}, 
            ylabel={Data size in KB}, 
            xmin=0, 
            xmax=200, 
            yticklabel style={yshift=0ex}, 
            enlarge y limits=0.2, 
            y=15pt, 
            grid=both, 
            legend style={at={(0.5,1.0)}, anchor=south, 
                          legend columns=2} 
        ]
        
        \addplot[fill=blue!50] coordinates {(89.05,64) (94.858,128) (109.296,256) (122.713,512)};
        \addplot[fill=green!50] coordinates {(11.007,64) (15.039,128) (25.023,256) (45.055,512)};

        
        \legend{OS, Transfer}
        
        \end{axis}
    \end{tikzpicture}
    \caption{Breakdown of UVM page transfer latency.  Note that host involvement overheads during the page fault are around 7$\times$ higher than the transfer time at 64KB page size.}
    \label{fig:latency_bar}
\end{figure}




%% file: system.tex

\mytitle{} provides an efficient GPU-managed paging system that automates the management of memory across a GPU and a host processor, while achieving high performance.  Programmers use a high-level programming abstraction to gain access to \mytitle{} to extend the available memory to GPUs and other accelerators. \mytitle{} uses an RNIC as a mediator of the requests between GPU threads and memory as shown in Figure ~\ref{fig:gpuvm_details}, since current systems do not support direct GPU-initiated RDMA access to host memory. 
In this context, the available extended memory can either be the CPU/host memory, remote memory of other nodes in the cloud, or memory of other GPUs or accelerators connected through an RDMA-capable network (in this paper, we only explore the first alternative). 
\input{rdma_diagram}

\mytitle{} needs to address a number of challenges: 
(1) \textbf{Supporting GPU-initiated memory transfers}: Systems are designed for the CPU to manage network-connected devices. To use the NIC for memory request handling, GPU threads need to have access to the control resources of the NICs. Additionally, we need to make sure \mytitle{} can efficiently generate parallel requests to be handled concurrently and efficiently by the RNIC.  
(2) \textbf{Supporting oversubscribed memory}: One of \mytitle{}'s goals is to provide efficient GPU memory oversubscription such that it can support workloads with larger memory requirements than physically available in GPU memory. Since memory is managed by the GPU, it must be able to evict pages to make room for newly fetched pages.  \mytitle{} must ideally map the pages in host memory to GPU memory which can possibly prevent early eviction of pages from GPU memory before they are used under memory oversubscription. 
(3) \textbf{Programming Abstraction}: Since the GPU threads are not typically designed to make memory accesses/requests by themselves, \mytitle{} needs to offer high-level abstractions that conceal its complexity and simplify the integration of \mytitle{} into existing GPU applications for programmers. 
In the remainder of this section, we first present an overview of \mytitle{}, and then discuss how we address these three challenges.

\subsection{System Overview}
\label{sec:system_overview}

We first overview the overall operation of \mytitle{}.  In systems such as UVM that rely on OS page fault handling~\cite{ziabari2016umh, ganguly2019interplay, kim2020batch, koukos2016building, choi2022memory, go2023early, markthub2018dragon}, the OS is responsible for memory allocation, controlling virtual-to-physical address translations, and page table update.
 \mytitle{} aims to shift the memory management to the GPU, removing the host OS from the critical path, and has to reimplement the memory management functionality on the GPU side.  Since current hardware does not allow the GPU to initiate RDMA transactions from CPU memory, \mytitle{} uses an RNIC to facilitate the transfer.  RNICs support one-sided RDMA connections to the CPU which allow a device to move pages from a remote device directly (without OS involvement).  

As shown in Figure ~\ref{fig:gpuvm_details}, \mytitle{} enables GPU threads to directly communicate with the RNIC by submitting work requests to the queue pair (QP). GPU threads are notified of the request completion through completion queue (CQ) entries inserted by the RNIC once a request has been serviced. \mytitle{} uses a memory region called GPUVM memory within device memory with device page tables and page maps efficiently managed by parallel GPU threads. The GPU hosts the GPUVM memory region, queue metadata, and parallel RDMA queues. 

\begin{figure}[t]
    \includegraphics[width=0.47\textwidth]{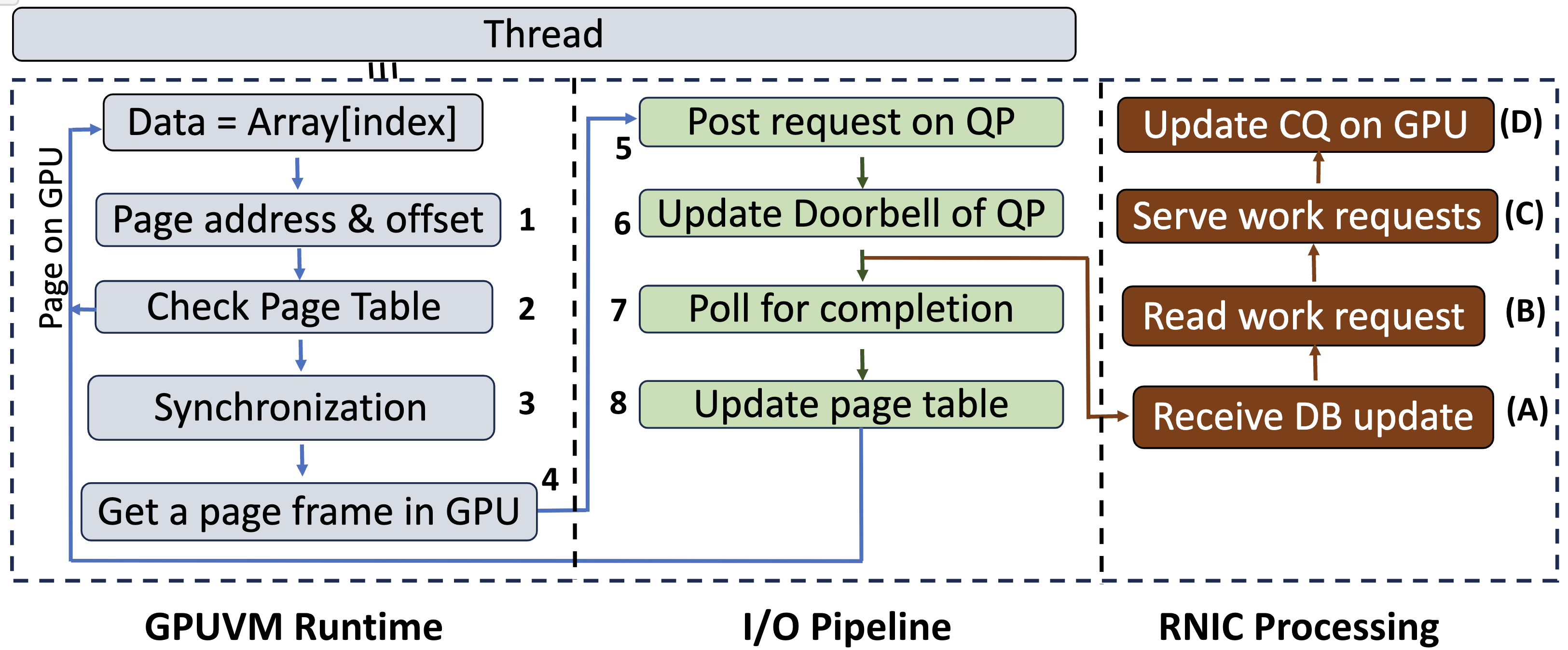}
    \caption{\mytitle{} system workflow for a single thread}
    \label{fig:system_overview}
    \vspace{-0.1in}
\end{figure}

 An important part of the abstraction is the \textit{\mytitle{}::class}  structure, which enables direct integration of \mytitle{} into existing GPU kernels.  We show an example in Listing ~\ref{lst:vector_addition} for a vector addition kernel; the code is minimally modified to include \mytitle{}'s array-like data structure.

Figure \ref{fig:system_overview} shows how a thread can access the page in \mytitle{}. First, the thread that accesses the \mytitle{}::class buffer at a specific index calculates the page address/number and the offset (1) and checks the page table for this page address (2). If the page is currently in the GPU memory, the thread can safely access the offset within the page. Otherwise, \mytitle{} runtime synchronizes all threads that access the same page (3).  
Within the synchronization block, a leader thread is selected to handle the page fault which starts with obtaining a page frame in GPU memory (4). 
Once a page frame is obtained, the leader thread prepares a work request for the page fault which includes a unique request number, page frame address, host memory address, remote key for host memory, and the QP ID. Once the leader thread inserts the request into the QP (5) and updates the doorbell register (6), it starts polling the CQ entry associated with the request number (7). More details in Section \ref{sec:paged_memory}.

On the RNIC side, when the doorbell update is received, the RNIC starts fetching the work requests from the QP in the GPU memory. The work requests for the page faults from the GPU threads are served by the RNIC. 
  For each page fault, the RNIC fetches the page from the CPU memory and sends it to the GPU memory (\mytitle{} memory in Figure ~\ref{fig:gpuvm_details}). Once the page is delivered to the assigned page frame, the RNIC updates the related CQ entry to notify the leader thread that the page fault has been serviced.

\subsection{Efficient RDMA I/O Management}
\label{sec:efficient_rdma}

To achieve high transfer throughput, \mytitle{} uses a memory-efficient streamlined I/O pipeline, which is shown in Figure ~\ref{fig:system_overview}. This pipeline enables GPU threads to manage page faults, post a request, and poll for completion.  The GPUVM runtime also has the GPU leader threads directly accessing RNIC I/O resources and updating page table entries. 
Sending a page fault request can be divided into two actions; inserting a work request for the page and ringing the doorbell which means writing the number of the request and index of the queue to the doorbell register. As the doorbell registers are located in the system memory, updating them for a single request would cause a serialization, thus, increasing the latency if the number of queues is not sufficient. 

The leader thread gets assigned a queue index that identifies which QP and CQ the leader will use for posting and polling. The leader atomically gets a number for the request called ~\textit{post\_number} that it uses as an ID of the page fault. Faults are handled in batches. If the post number is less than the fault batch, then the leader can continue to insert the request into the send queue with ~\textit{post\_number}. Otherwise, it must wait for the current batch to finish. For a fault batch, several leader threads insert requests for different pages. However, only one randomly selected leader (the one that atomically gets the lock of the queue) updates the doorbell. Once the queue is locked no other request can be inserted. 

\input{listing_addvector}

To make sure all the leaders within the batch finished inserting into the send queue\footnote{Queue pair (QP) contains 2 queues; send and receive queues. For \mytitle{}, only send queue is utilized.}, all the threads atomically increment a variable called ~\textit{batch\_counter}. If this variable matches with the max post number, then all the leaders have finished the insertion of page requests into the send queue. At this stage, the leader with the queue lock can update the doorbell. After the update of the doorbell, all the leader threads in the batch start polling on completion queue entries corresponding to ~\textit{post\_number}.

The period during which leader threads wait for completion is consistent, depending on the NIC, and threads release the lock once their requests are fulfilled, leaving no opportunity for security channels or performance slowdown attacks.

The queue depth to achieve the max available throughput is covered by  Little's law ~\cite{little2008little}, $L=\lambda*W$, where $L$, $\lambda$, and $W$ represent the average queue depth, access latency, and target throughput, respectively. The target throughput is $12GBps$ as we are using a PCIe-3-based machine. The latency $\lambda$ is $23\mu s$ empirically estimated on the testbed we use in our experiments.  From Little's law, we obtain an average queue depth of $36$ for an 8KB page and $72$ for a 4KB page. Therefore, optimal performance can be achieved with a batch of one fault with more than 72$(23\mu *12GBps/4KB)$ parallel queues (for 4KB pages and, similarly, 36 queues for 8KB pages).

\input{eviction}

\subsection{Paged Memory System Design}
\label{sec:paged_memory}

Systems that rely on OS virtual memory for data migration employ two important steps for acquiring the page on demand; memory allocation and address translation which happen during application runtime as shown in Figure \ref{fig:uvm_details}. These steps involve critical sections with severe serialization. However, \mytitle{}, relying on RNIC hardware for address translation, can complete many more requests in flight and parallel, bypassing the host OS. As the memory is already allocated before the application starts running with \mytitle{}, no OS kernel involvement is needed during the execution of the application. Implementation details in Section \ref{sec:implementation}.  

Within the \mytitle{} runtime, the host virtual memory can be considered as "Physical Address Space" and GPU virtual memory as "Virtual Address Space" as shown in Figure \ref{fig:virtualGPU}. The host memory contains all the application data. GPU memory can host all the pages and sometimes becomes limited as the workload sizes get bigger than GPU memory.

~\textbf{Page mapping}. The main goal of \mytitle{} runtime is to dynamically and efficiently map the pages from 'physical' to `virtual' address space which can be defined as mapping the virtual page on host memory to a virtual page on GPU memory as shown in Figure ~\ref{fig:virtualGPU}. 
Upon a request from GPU threads, RNIC hardware brings the page to the virtual pages on GPU memory. After all the threads complete the read/write operation on the page, it can be safely evicted if necessary. The mapping of pages is stored in GPU memory for fast access and dynamically modified by GPU threads during application execution.

\input{thread_coalescing}

~\textbf{Eviction scheme.} The \mytitle{} implementation specifically aims to improve the ease of programming for developers and eliminate the need for partitioning of the workload and the manual transfers of the partitions that are needed when the workload size is bigger than the available GPU memory. To achieve stable performance as the oversubscription level increases, the efficient FIFO page eviction mechanism is implemented. 

The GPU virtual memory can be viewed as a circular queue (ring buffer) with a global head cursor depicted in Figure ~\ref{fig:virtualGPU}. Each entry in this queue represents a page mapping. When the leader thread handles the page fault, a new page mapping that maps the host page to the GPU page is established. The leader thread is responsible for establishing the mapping through \mytitle{} runtime library. It should be noted leader thread atomically gets the mapping and is not allowed to map arbitrarily to any GPU page frame. After mapping is established, the leader thread checks if the page frame is mapped to another page. If the page frame is already mapped to another page, it waits for that page to be released. This happens through checking \textbf{reference counter} which shows the number of warps that currently access/need the page. Once the reference counter becomes zero, the leader thread immediately evicts the page to host and creates the new mapping.


\sloppy
~\textbf{Page access synchronization.} We use opportunistic warp level synchronization primitive ~\cite{nvidia2021warp}, \texttt{\_\_match\_any\_sync}, to select a leader within warp threads with the same \texttt{\_\_activemask}. \mytitle{} runtime also incorporates inter-warp coalescing, considering that there might be more than one warp that access the same page; within all the leaders from those warps, only one leader will lock the page entry as illustrated in Figure ~\ref{fig:thread_coalescing}. Page fault in Figure ~\ref{fig:thread_coalescing} involves getting a mapping, triggering a work request on RNIC, and eviction of another page if necessary. Once the page fault is complete, all threads can continue processing the data. 


\input{topology}

\subsection{Comparisons with UVM and Bulk Transfer}
Unified virtual memory makes the host and GPU memories virtually connected allowing the device code to directly access the systems memory provide programmer-agnostic development experience. The bulk transfer involves partitioning the data and transferring the partitions to GPU memory by CPU. There are \textbf{several clear advantages of \mytitle{} over UVM and bulk transfer approaches}. 

Firstly, UVM transfer size is 4KB (for $x86\_64$) and 60KB is asynchronously transferred due to speculative prefetching to complete the size to GPU page size and boost the performance
~\cite{allen2024fine, schieffer2024harnessing}. The eviction size is 2MB ~\cite{allen2024fine} which is called a virtual address block (VABlock). This can sometimes be the bottleneck for oversubscription as the newly fetched page can be evicted within VABlock. However, since in \mytitle{}, GPU threads have direct access to the page table and can monitor the page status and reference counter, eviction becomes much more efficient. For example, if the page is write-intensive, \mytitle{} can delay its eviction and evict one of the least needed read-intensive pages.

Secondly, to optimize the UVM, the application developer should statically specify the access hints before the application starts running on GPU. Thus, UVM cannot leverage dynamic memory optimizations, making it more developer-dependent. On the other hand, \mytitle{} can benefit from dynamic and efficient page mapping optimizations such as reference counters for each page.

Furthermore, the commonly used method is partitioning the dataset and loading the partitions into GPU memory for computation. In this way, the application's performance can benefit from the GPU's high memory bandwidth. However, this method requires the data and algorithm to be actively partitioned/tiled by the developer. For data-dependent applications such as graph workloads, recommender systems, data analytics, etc., it might be very difficult to find an ideal partitioning algorithm. To this end, \mytitle{} becomes an advantageous alternative providing on-demand access to the data.


%% file: rdma_diagram.tex
\begin{figure}[t]
    \begin{tikzpicture}

        \tikzset{
            sm/.style={rectangle, rounded corners, draw, text centered, minimum width=3cm, minimum height=1.5cm, fill=blue!20},
            smallrect/.style={rectangle, draw, minimum width=2.5cm, minimum height=1cm, fill=green!20},
            cpu/.style={rectangle, rounded corners, draw, text centered, minimum width=3cm, minimum height=3cm}, 
            smallnode/.style={rectangle, draw, minimum width=1cm, minimum height=1cm, fill=orange!20, text centered}, 
            node2/.style=node2/.style={rectangle, draw, fill=gray!50, minimum width=1.5cm, minimum height=0.8cm, rotate=90, text centered}, 
            ->/.style={draw, -{Stealth[length=1.5mm, width=1mm]}},
            colored/.style={fill=yellow!100},
            colored2/.style={fill=green!30}
        }

        \node (sm) [sm, minimum height=0cm, minimum width=0cm, rotate=90, fill=white, draw=none] {};

        \node (sm1) [sm, minimum height=1.2cm, minimum width=1.0cm, yshift=-1.2cm, xshift=-0.1cm, rotate=90] {\tiny gpuvm[tid]};
        \node[rotate=90] at (-0.5, -1.2) {\tiny \textbf{PU\#0}}; 

        \node[rotate=90, right of=sm, xshift=-1.0cm] {\huge $\cdots$}; 

        \node (sm2) [sm, minimum height=1.2cm, minimum width=1.0cm, yshift=1.0cm, xshift=-0.1cm, rotate=90] {\tiny Data[tid]};
        \node[rotate=90] at (-0.5, 1) {\tiny \textbf{PU\#N}}; 



        \node (devicememory) [sm, right of=sm, yshift=0.0cm, xshift=0.8cm, minimum height=3.6cm, minimum width=2.1cm, fill=blue!3, text=black] {\tiny \parbox[c]{1.8cm}{\centering \vspace{-3.1cm} Device Memory}};

        \node (restMem) [sm, right of=devicememory, xshift=-1.0cm, yshift=1.25cm, minimum height=0.4cm, minimum width=2.0cm, line width=0.0mm, fill=blue!10, text=black] {\tiny \parbox[c]{1.76cm}{\centering Rest of the memory}};

        \node (rdmaMem) [sm, right of=devicememory, xshift=-1.0cm, yshift=0.60cm, minimum height=0.8cm, minimum width=2.0cm, line width=0.0mm, fill=blue!10, text=black] {\tiny \parbox[c]{1.76cm}
        {\centering \mytitle{} Memory}};

        \node (gpuQueues) [sm, right of=devicememory, xshift=-1.0cm, yshift=-0.80cm, minimum height=1.9cm, minimum width=2.0cm, line width=0.0mm, fill=blue!10, text=white] 
        {
            \tiny \parbox[c]{1.76cm}{
                \centering 
                \vspace{-1.2cm} 
                \tikz[baseline=(content.base)] 
                \node[draw=black, minimum width=1.8cm, minimum height=0.4cm, inner sep=2pt, fill=blue!70] (content) { Queue metadata};
            }
        };

        \node (queuesWholeBlock) [sm, right of=gpuQueues, xshift=-0.95cm, yshift=-0.25cm, minimum height=1.30cm, minimum width=1.8cm, line width=0.0mm, fill=gray!5, text=black]{};

        \node (queueBlock1) [sm, right of=gpuQueues, xshift=-1.4cm, yshift=-0.25cm, minimum height=1.2cm, minimum width=0.65cm, line width=0.0mm, fill=gray!25, text=black]{};

        \node (queueBlock2) [sm, right of=gpuQueues, xshift=-0.45cm, yshift=-0.25cm, minimum height=1.2cm, minimum width=0.65cm, line width=0.0mm, fill=gray!25, text=black]{};

        \node[rotate=90, right of=gpuQueues, xshift=-1.2cm, yshift=-0.1cm] { $\cdots$};

        \def\cX{2.35}
        \def\cY{-0.77}
        \def\R1{0.2}
        \def\R2{0.28}
        \draw[] (\cX, \cY) circle (0.28cm);
        
        \draw[] (\cX, \cY) circle (0.2cm);
        
            \foreach \angle in {0, 24, 48, 72, 96, 120, 144, 168, 192, 216, 240, 264, 288, 312, 336} {
                \draw[] ({\cX + 0.2*cos(\angle)},{\cY + 0.2*sin(\angle)}) -- ({\cX + 0.28*cos(\angle)},{\cY + 0.28*sin(\angle)});
            }
        
        \def\base{3}
        \foreach \i in {6, 7, 8, 9, 10, 11, 12} {
            \pgfmathsetmacro{\start}{\base + 24 * \i}
            \pgfmathsetmacro{\endA}{\start + 19}
    
            \fill[colored] ({\cX + 0.21*cos(\start)},{\cY + 0.21*sin(\start)}) 
                arc[start angle=\start, end angle=\endA, radius=0.2cm] --
                ({\cX + 0.27*cos(\endA)},{\cY + 0.27*sin(\endA)}) 
                arc[start angle=\endA, end angle=\start, radius=0.28cm] -- cycle;
        }
        \pgfmathsetmacro{\start}{\base + 24 * 5}
            \pgfmathsetmacro{\endA}{\start + 19}

        \fill[colored2] ({\cX + 0.21*cos(\start)},{\cY + 0.21*sin(\start)}) 
                arc[start angle=\start, end angle=\endA, radius=0.2cm] --
                ({\cX + 0.27*cos(\endA)},{\cY + 0.27*sin(\endA)}) 
                arc[start angle=\endA, end angle=\start, radius=0.28cm] -- cycle;
        \node at (\cX, \cY) {\tiny \textbf{QP}};

        \def\cX{1.4}
        \def\cY{-0.77}
        \def\R1{0.2}
        \def\R2{0.28}
        \draw[] (\cX, \cY) circle (0.28cm);
        
        \draw[] (\cX, \cY) circle (0.2cm);
        
            \foreach \angle in {0, 24, 48, 72, 96, 120, 144, 168, 192, 216, 240, 264, 288, 312, 336} {
                \draw[] ({\cX + 0.2*cos(\angle)},{\cY + 0.2*sin(\angle)}) -- ({\cX + 0.28*cos(\angle)},{\cY + 0.28*sin(\angle)});
            }
        
        \def\base{3}
        \foreach \i in {6, 7, 8, 9, 10, 11, 12} {
            \pgfmathsetmacro{\start}{\base + 24 * \i}
            \pgfmathsetmacro{\endA}{\start + 19}
    
            \fill[colored] ({\cX + 0.21*cos(\start)},{\cY + 0.21*sin(\start)}) 
                arc[start angle=\start, end angle=\endA, radius=0.2cm] --
                ({\cX + 0.27*cos(\endA)},{\cY + 0.27*sin(\endA)}) 
                arc[start angle=\endA, end angle=\start, radius=0.28cm] -- cycle;
        }
        \pgfmathsetmacro{\start}{\base + 24 * 5}
            \pgfmathsetmacro{\endA}{\start + 19}

        \fill[colored2] ({\cX + 0.21*cos(\start)},{\cY + 0.21*sin(\start)}) 
                arc[start angle=\start, end angle=\endA, radius=0.2cm] --
                ({\cX + 0.27*cos(\endA)},{\cY + 0.27*sin(\endA)}) 
                arc[start angle=\endA, end angle=\start, radius=0.28cm] -- cycle;
        \node at (\cX, \cY) {\tiny \textbf{QP}};

        \def\cX{1.4}
        \def\cY{-1.35}
        \def\R1{0.2}
        \def\R2{0.28}
        \draw[] (\cX, \cY) circle (0.28cm);
        
        \draw[] (\cX, \cY) circle (0.2cm);
        
            \foreach \angle in {0, 24, 48, 72, 96, 120, 144, 168, 192, 216, 240, 264, 288, 312, 336} {
                \draw[] ({\cX + 0.2*cos(\angle)},{\cY + 0.2*sin(\angle)}) -- ({\cX + 0.28*cos(\angle)},{\cY + 0.28*sin(\angle)});
            }
        
        \def\base{3}
        \foreach \i in {5, 6, 7, 8, 9, 10, 11} {
            \pgfmathsetmacro{\start}{\base + 24 * \i}
            \pgfmathsetmacro{\endA}{\start + 19}
    
            \fill[colored] ({\cX + 0.21*cos(\start)},{\cY + 0.21*sin(\start)}) 
                arc[start angle=\start, end angle=\endA, radius=0.2cm] --
                ({\cX + 0.27*cos(\endA)},{\cY + 0.27*sin(\endA)}) 
                arc[start angle=\endA, end angle=\start, radius=0.28cm] -- cycle;
        }
        \pgfmathsetmacro{\start}{\base + 24 * 12}
            \pgfmathsetmacro{\endA}{\start + 19}

        \fill[colored2] ({\cX + 0.21*cos(\start)},{\cY + 0.21*sin(\start)}) 
                arc[start angle=\start, end angle=\endA, radius=0.2cm] --
                ({\cX + 0.27*cos(\endA)},{\cY + 0.27*sin(\endA)}) 
                arc[start angle=\endA, end angle=\start, radius=0.28cm] -- cycle;
        \node at (\cX, \cY) {\tiny \textbf{CQ}};

        \def\cX{2.35}
        \def\cY{-1.35}
        \def\R1{0.2}
        \def\R2{0.28}
        \draw[] (\cX, \cY) circle (0.28cm);
        
        \draw[] (\cX, \cY) circle (0.2cm);
        
            \foreach \angle in {0, 24, 48, 72, 96, 120, 144, 168, 192, 216, 240, 264, 288, 312, 336} {
                \draw[] ({\cX + 0.2*cos(\angle)},{\cY + 0.2*sin(\angle)}) -- ({\cX + 0.28*cos(\angle)},{\cY + 0.28*sin(\angle)});
            }
        
        \def\base{3}
        \foreach \i in {5, 6, 7, 8, 9, 10, 11} {
            \pgfmathsetmacro{\start}{\base + 24 * \i}
            \pgfmathsetmacro{\endA}{\start + 19}
    
            \fill[colored] ({\cX + 0.21*cos(\start)},{\cY + 0.21*sin(\start)}) 
                arc[start angle=\start, end angle=\endA, radius=0.2cm] --
                ({\cX + 0.27*cos(\endA)},{\cY + 0.27*sin(\endA)}) 
                arc[start angle=\endA, end angle=\start, radius=0.28cm] -- cycle;
        }
        \pgfmathsetmacro{\start}{\base + 24 * 12}
            \pgfmathsetmacro{\endA}{\start + 19}

        \fill[colored2] ({\cX + 0.21*cos(\start)},{\cY + 0.21*sin(\start)}) 
                arc[start angle=\start, end angle=\endA, radius=0.2cm] --
                ({\cX + 0.27*cos(\endA)},{\cY + 0.27*sin(\endA)}) 
                arc[start angle=\endA, end angle=\start, radius=0.28cm] -- cycle;
        \node at (\cX, \cY) {\tiny \textbf{CQ}};

        \node (pcie1) [sm, right of=sm, xshift=2.4cm, yshift=0.0cm, minimum height=0.2cm, minimum width=3.8cm, rotate=90, fill=lightgray, dashed] {\tiny PCIe Interconnect};

        \node (rnic1) [sm, right of=pcie1, xshift=0.3cm, yshift=1.01cm, minimum height=2.0cm, minimum width=1.7cm] {\tiny \parbox[c]{0.8cm}{\centering \vspace{-1.6cm} RNIC\#1}};

        \node (rnic1Mem) [sm, right of=rnic1, xshift=-1.0cm, yshift=0.4cm, minimum height=0.2cm, minimum width=1.6cm, line width=0.0mm, fill=blue!10, text=black] {\tiny \parbox[c]{0.8cm}{\centering On-chip buffers}};

        
        \node (rnic1ctrl) [sm, right of=rnic1, xshift=-1.00cm, yshift=-0.48cm, minimum height=0.95cm, minimum width=1.6cm, line width=0.0mm, fill=white, text=black] {};
        
        \node at ($(rnic1ctrl.east) + (-0.15cm,-0.0cm)$) [rotate=90, fill=none, text=black, anchor=center] {\tiny CTRL};

        \node (rnic1DMA) [sm, right of=rnic1, xshift=-1.15cm, yshift=-0.3cm, minimum height=0.2cm, minimum width=1.0cm, line width=0.0mm, fill=blue!10, text=black] {\tiny \parbox[c]{0.98cm}{\centering DMA}};

        \node (rnic1db) [sm, right of=rnic1, xshift=-1.15cm, yshift=-0.7cm, minimum height=0.2cm, minimum width=1.0cm, line width=0.0mm, fill=blue!10, text=black] {\tiny \parbox[c]{0.98cm}{\centering DB REG \#N}};

        \node (rnic2) [sm, right of=pcie1, xshift=0.3cm, yshift=-1.01cm, minimum height=2.0cm, minimum width=1.7cm] {\tiny \parbox[c]{0.8cm}{\centering \vspace{-1.6cm} RNIC\#2}};



        \node (rnic2Mem) [sm, right of=rnic2, xshift=-1.0cm, yshift=0.4cm, minimum height=0.2cm, minimum width=1.6cm, line width=0.0mm, fill=blue!10, text=black] {\tiny \parbox[c]{0.8cm}{\centering On-chip buffers}};

        
        \node (rnic2ctrl) [sm, right of=rnic2, xshift=-1.00cm, yshift=-0.48cm, minimum height=0.95cm, minimum width=1.6cm, line width=0.0mm, fill=white, text=black] {};
        
        \node at ($(rnic2ctrl.east) + (-0.15cm,-0.00cm)$) [rotate=90, fill=none, text=black, anchor=center] {\tiny CTRL};

        \node (rnic2DMA) [sm, right of=rnic2, xshift=-1.15cm, yshift=-0.3cm, minimum height=0.2cm, minimum width=1.0cm, line width=0.0mm, fill=blue!10, text=black] {\tiny \parbox[c]{0.98cm}{\centering DMA}};

        \node (rnic2db) [sm, right of=rnic2, xshift=-1.15cm, yshift=-0.7cm, minimum height=0.2cm, minimum width=1.0cm, line width=0.0mm, fill=blue!10, text=black] {\tiny \parbox[c]{0.98cm}{\centering DB REG \#N}};


        \node (gpu) [cpu, right of=sm, xshift=0.1cm, minimum height=3.8cm, minimum width=3.7cm] {};



        \node (pcie2) [sm, right of=pcie1, xshift=1.5cm, yshift=0.0cm, minimum height=0.2cm, minimum width=3.8cm, rotate=90, fill=lightgray, dashed] {\tiny PCIe Interconnect};

        \node (hostmemory) [sm, right of=pcie2, xshift=-0.1 cm, yshift=0cm, minimum height=4.0cm, minimum width=1.15cm, fill=brown!40, text=black] {\tiny \parbox[c]{0.8cm}{\centering \vspace{-3.3cm}Host Memory}};

        \node (memBlock1) [sm, right of=hostmemory, xshift=-1.0cm, yshift=0.8cm, minimum height=1.2cm, minimum width=1.0cm, line width=0.0mm, fill=white, text=black] {};

        \node (memoryDMA1) [sm, right of=hostmemory, xshift=-1.25cm, yshift=0.8cm, minimum height=0.2cm, minimum width=0.5cm, line width=0.0mm, fill=blue!10, text=black, rotate=90] {\tiny \parbox[c]{0.8cm}{\centering DMA}};

        \draw[<->, double, double distance=2.0pt, shorten <=1pt, shorten >=1pt, >=Stealth] 
          ($(memoryDMA1.north)+ (0.15, 0.0)$) -- ($(pcie2.south) + (-0.2, 0.785)$) 
          node[midway, above] {};

        \node (memoryContent1) [sm, right of=hostmemory, xshift=-0.8cm, yshift=0.8cm, minimum height=0.2cm, minimum width=0.5cm, line width=0.0mm, fill=blue!10, text=black, rotate=90] {\tiny \parbox[c]{0.8cm}{\centering Content}};

        \node[rotate=90, right of=hostmemory, xshift=-1.2cm] {\huge $\cdots$}; 

        \node (memBlock2) [sm, right of=hostmemory, xshift=-1.0cm, yshift=-1.3cm, minimum height=1.2cm, minimum width=1.0cm, line width=0.0mm, fill=white, text=black] {};

        \node (memoryDMA2) [sm, right of=hostmemory, xshift=-1.25cm, yshift=-1.3cm, minimum height=0.2cm, minimum width=0.5cm, line width=0.0mm, fill=blue!10, text=black, rotate=90] {\tiny \parbox[c]{0.8cm}{\centering DMA}};

        \draw[<->, double, double distance=2.0pt, shorten <=1pt, shorten >=1pt, >=Stealth] 
          ($(memoryDMA2.north)+ (0.15, 0.0)$) -- ($(pcie2.south) + (-0.2, -1.3)$) 
          node[midway, above] {};

        \node (memoryContent1) [sm, right of=hostmemory, xshift=-0.8cm, yshift=-1.3cm, minimum height=0.2cm, minimum width=0.5cm, line width=0.0mm, fill=blue!10, text=black, rotate=90] {\tiny \parbox[c]{0.8cm}{\centering Content}};

        \node at (0.0, 1.8) {\tiny \textbf{GPU}}; 

        \draw[<->, double, double distance=2.0pt, shorten <=1pt, shorten >=1pt, >=Stealth] 
          ($(rnic1Mem.east)+ (-0.3, 0.0)$) -- ($(pcie2.north) + (0.2, 1.4)$) 
          node[midway, above] {};

        \draw[<->, double, double distance=2.0pt, shorten <=1pt, shorten >=1pt, >=Stealth] 
          ($(rnic2Mem.east)+ (-0.3, 0.0)$) -- ($(pcie2.north) + (0.2, -0.6)$) 
          node[midway, above] {};

        \draw[<->, double, double distance=2.0pt, shorten <=1pt, shorten >=1pt, >=Stealth] 
          ($(rnic2DMA.west)+ (0.1, 0.0)$) -- ($(pcie1.south) + (-0.2, -1.3)$) 
          node[midway, above] {};

        \draw[<->, double, double distance=2.0pt, shorten <=1pt, shorten >=1pt, >=Stealth] 
          ($(rnic1DMA.west)+ (0.1, 0.0)$) -- ($(pcie1.south) + (-0.2, 0.7)$) 
          node[midway, above] {};

        \draw[<->, double, double distance=2.0pt, shorten <=1pt, shorten >=1pt, >=Stealth] 
          ($(rdmaMem.east)+ (-0.1, 0.0)$) -- ($(pcie1.north) + (0.2, 0.60)$) 
          node[midway, above] {};

        \draw[->, dashed] ($(gpuQueues.east)+ (0.0, 0.2)$) -- ++(0, 0) -- ++(0.3, 0) -- ++(0, 0.90) -- (rnic1db.west);

        \draw[->, dashed] ($(gpuQueues.east)+ (0.0, -0.2)$) -- ++(0, 0) -- ++(0.3, 0) -- ++(0, -0.70) -- (rnic2db.west);

        \draw[->] ($(sm1.south)+ (0.0, 0.0)$) -- ++(0, 0) -- ++(0.1, 0) -- ++(0, 1.8) -- (rdmaMem.west);

        \draw[->] ($(sm2.south)+ (0.0, 0.0)$) -- ++(0, 0) -- ++(0.1, 0) -- ++(0, 0.25) -- (restMem.west);
        

        \draw[<->, double, double distance=2.0pt, shorten <=1pt, shorten >=1pt, >=Stealth] 
          ($ (-0.8, -2.3)$) -- ($(0.0, -2.3)$) 
          node[midway, above] {};

        \node at (0.5, -2.3) {\tiny \textbf{Data Path}};

        \draw[->, dashed] 
          ($ (1.5, -2.3)$) -- ($(2.2, -2.3)$) 
          node[midway, above] {};

        \node at (2.9, -2.3) {\tiny \textbf{Control Path}};

        \draw[->] 
          ($ (4.0, -2.3)$) -- ($(4.6, -2.3)$) 
          node[midway, above] {};

        \node at (5.6, -2.3) {\tiny \textbf{GPU memory access}};

    \end{tikzpicture}
    \caption{Schematic representation of GPUVM design.}
    \label{fig:gpuvm_details}
\end{figure}

%% file: listing_addvector.tex
\lstset{
    basicstyle=\tiny\ttfamily, 
    columns=flexible,           
    frame=single,               
    breaklines=true,            
    linewidth=0.47\textwidth    
}

\begin{lstlisting}[language=C++, caption=Vector addition with \mytitle{}, label=lst:vector_addition, float=t]
__global__ 
void vectorAdd(gpuvm<float> A, gpuvm<float> B, gpuvm<float> C, int N) {
    int i = blockDim.x * blockIdx.x + threadIdx.x;
    if (i < N) {
        C[i] = A[i] + B[i];
    }
}
\end{lstlisting}

%% file: eviction.tex
\begin{figure}[t]
\resizebox{0.43\textwidth}{!}{
\begin{tikzpicture}[
    every node/.style={font=\sffamily, scale=1},
    page/.style={draw, rectangle, minimum width=2cm, minimum height=0.5cm},
    arrow/.style={-Latex, thick, blue},
    wavyborder/.style={decorate, decoration={snake, amplitude=0.2cm, segment length=0.5cm}, draw, thick, red},
    cloud/.style={draw, cloud, cloud puffs=10, cloud puff arc=120, aspect=2, thick, fill=blue!20, minimum width=3cm, minimum height=1.2cm},
    colored/.style={fill=red!30},
    colored2/.style={fill=green!30}
]

    \node[page, align=center, colored] (page0) at (-3, 0) {Page 0};
    \node[page, align=center, colored2] (page1) at (-3, -0.55) {Page 1};
    \node[page, align=center, colored2] (page2) at (-3, -1.1) {Page 2};
    \node[page, align=center, colored2] (page3) at (-3, -1.65) {Page 3};
    \node[page, align=center, colored] (page4) at (-3, -2.2) {Page 4};
    \node[page, align=center, colored2] (page5) at (-3, -2.755) {Page 5};
    \node[page, align=center, minimum height=2cm] (dots) at (-3, -4.02) {\vdots};
    \node[page, align=center, colored] (pagen) at (-3, -5.3) {Page n};
    \node[align=center, scale=1.0, below=0.3cm of pagen] {Host memory};

    \node[page, align=center, colored] (vpage0) at (0, -1.5) {Page 0};
    \node[page, align=center, colored] (vpage1) at (0, -2.05) {Page 1};
    \node[page, align=center, colored2] (vpage2) at (0, -2.6) {Page 2};
    \node[page, align=center, minimum height=1cm] (vdots) at (0, -3.38) {\vdots};
    \node[page, align=center, colored] (vpagen) at (0, -4.15) {Page m};
    \node[align=center, scale=1.0, below=0.3cm of vpagen, xshift=0.5cm] {GPU memory};

    \def\cX{2.5}
    \def\cY{-2.7}
    \def\R1{0.7}
    \def\R2{1}
    \draw[ultra thick] (\cX, \cY) circle (1cm);
    
    \draw[ultra thick] (\cX, \cY) circle (0.7cm);
    
    \foreach \angle in {0, 15, 30, 45, 60, 75, 90, 105, 120, 135, 150, 165, 180, 195, 210, 225, 240, 255, 270, 285, 300, 315, 330, 345} {
        \draw[ultra thick] ({\cX + 0.7*cos(\angle)},{\cY + 0.7*sin(\angle)}) -- ({\cX + 1*cos(\angle)},{\cY + 1*sin(\angle)});
    }

    \def\base{17}

    \foreach \i in {5, 6, 7, 8, 9, 10, 11, 12, 13, 14, 15, 16, 17, 18, 19, 20, 21, 22} {
        \pgfmathsetmacro{\start}{\base + 15 * \i}
        \pgfmathsetmacro{\endA}{\start + 12}

        \fill[colored] ({\cX + 0.73*cos(\start)},{\cY + 0.73*sin(\start)}) 
            arc[start angle=\start, end angle=\endA, radius=0.7cm] --
            ({\cX + 0.97*cos(\endA)},{\cY + 0.97*sin(\endA)}) 
            arc[start angle=\endA, end angle=\start, radius=1cm] -- cycle;
    }

    \foreach \i in {0, 1, 2, 3, 4, 23, 24} {
        \pgfmathsetmacro{\start}{\base + 15 * \i}
        \pgfmathsetmacro{\endA}{\start + 12}

        \fill[colored2] ({\cX + 0.73*cos(\start)},{\cY + 0.73*sin(\start)}) 
            arc[start angle=\start, end angle=\endA, radius=0.7cm] --
            ({\cX + 0.97*cos(\endA)},{\cY + 0.97*sin(\endA)}) 
            arc[start angle=\endA, end angle=\start, radius=1cm] -- cycle;
    }


    \draw[arrow] (page0.east) -- (vpage0.west);
    \draw[arrow] (page4.east) -- (vpage1.west);
    \draw[arrow] (pagen.east) -- (vpagen.west);





    \draw[arrow]  (\cX + 1.2, \cY + 1.0) -- (\cX + 0.9, \cY + 0.1);

    \node[align=center, right=0.2cm] at (\cX + 0.0, \cY + 1.5) {Next available\\page};

    \draw[thick, blue] ({vpage0.east}) ++(0, 0.275) -- (\cX, \cY+1); 
    \draw[thick, blue] (vpagen.east) ++(0, -0.275) -- (\cX , \cY-1); 

\end{tikzpicture}
}
\caption{\mytitle{} page mapping: Host memory contains all pages. GPU memory is organized as a circular page buffer. Red represents mapped pages, and green unmapped pages.}
    \label{fig:virtualGPU}
\end{figure}
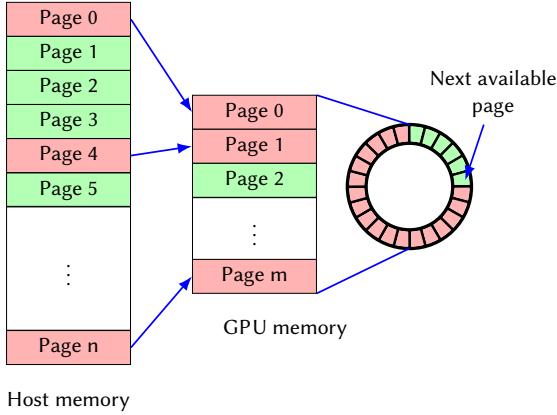


%% file: thread_coalescing.tex
\begin{figure}[t]
    \includegraphics[width=8.0cm]{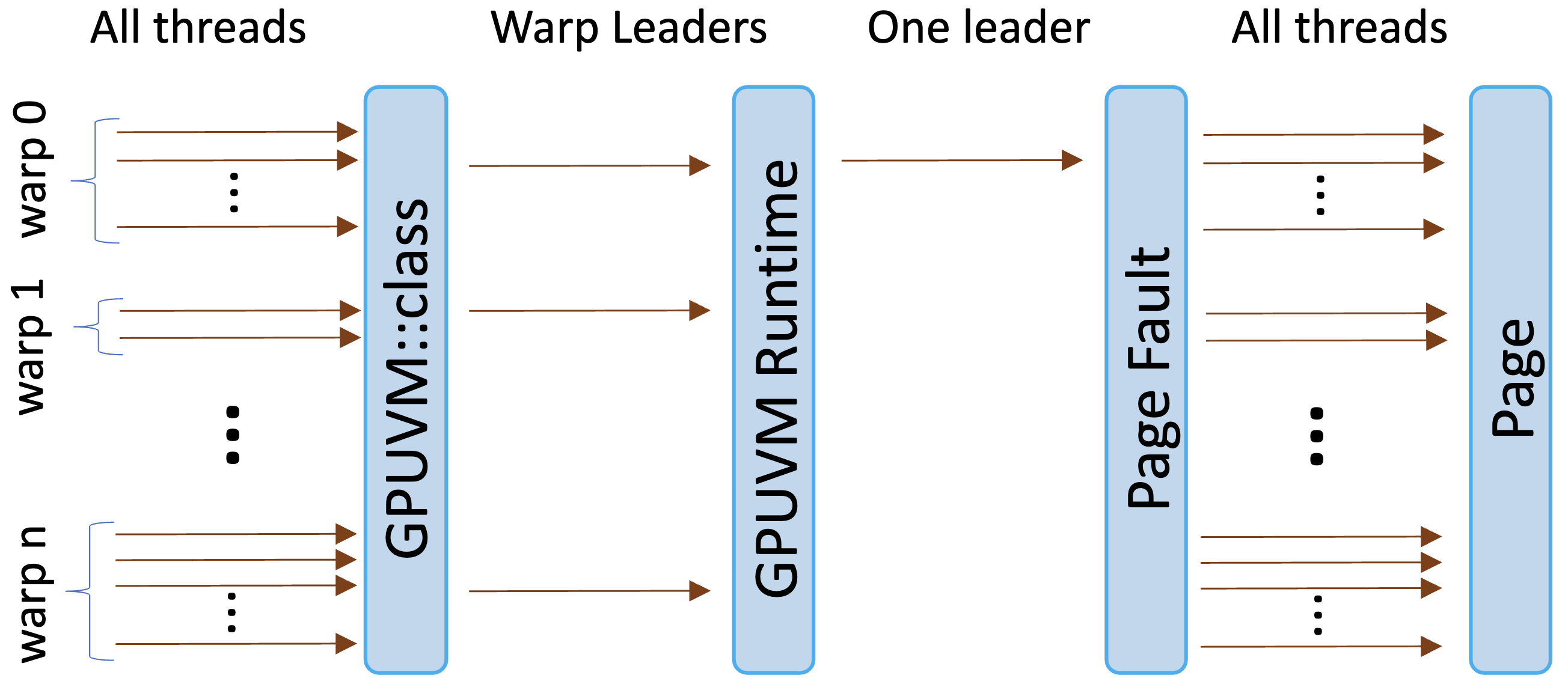}
    \caption{\mytitle{} runtime workflow. Threads access \mytitle{}::class and warp leaders are forwarded to \mytitle{} runtime. Page fault involves acquiring a mapping and posting and polling for a page request.}
    \label{fig:thread_coalescing}
\end{figure}

%% file: topology.tex
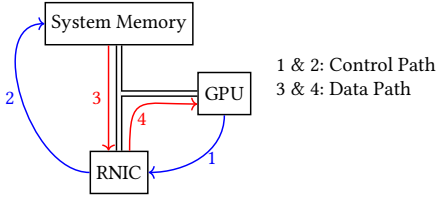
\begin{figure}[t]
    \raggedright
    \resizebox{0.35\textwidth}{!}{ 
    \begin{tikzpicture}[
        every node/.style={scale=1, font=\large}, 
        node distance=0.5cm and 1.8cm,
        box/.style={draw, minimum width=2cm, minimum height=0.8cm, text centered, thick},
        connection/.style={thick}
    ]

    \node (memorycontroller) [box] {System Memory};
    
    \node (nic0) [box, below=of memorycontroller, yshift=-1.5cm, xshift=0cm, minimum width=1cm] {RNIC};
    \node (gpu1) [box, below=of memorycontroller, yshift=0.0cm, xshift=2cm, minimum width=1cm] {GPU};

    \draw[thick, double, double distance=2.0pt] (memorycontroller.south) -- (nic0.north);

    \draw[thick, double, double distance=2.0pt] (gpu1.west) -- ++(-1.45cm, 0);

    \draw[thick, ->, blue] (gpu1.south) to[out=-90, in=0] 
        node[midway, right] {1} 
        (nic0.east);

    \draw[thick, ->, blue] (nic0.west) to[out=180, in=180] 
        node[midway, left] {2} 
        (memorycontroller.west);
    
    \draw[thick, ->, red] (memorycontroller.south) ++(-0.2cm, 0) -- ++(0, -2.0cm) 
        node[midway, left] {3}; 
    
    \draw[thick, ->, red] (nic0.north) ++(0.2cm, 0) to[out=90, in=180, looseness=1.8] 
        node[midway, below] {4} 
        ($(gpu1.west) + (0, -0.2cm)$); 

    \node[align=left] at (4.5, -1) {1 \& 2: Control Path\\3 \& 4: Data Path}; 
    \end{tikzpicture}
    }
    \caption{Configuration of r7525 node in Cloudlab ~\cite{Duplyakin+:ATC19}. As 3 \& 4 share the same PCIe bridge connecting NIC to the system, the available bandwidth drops to 8GBps, halving PCIe 3 bandwidth.}
    \label{fig:system_topology}
    \vspace{-10pt}
\end{figure}

%% file: table_hardware.tex
\begin{flushleft} 

\begin{table}[t] 
\caption{System configuration for experiments}
\label{tab:table_hardware}
\resizebox{0.47\textwidth}{!}{
\begin{tabular}{|l|l|}
\hline
\textbf{Component}        & \textbf{Specification}                              \\ \hline
\textbf{CPU}              & 2$\times$ AMD 7542 (32 cores, 2.40 GHz)          
\\ \hline
\textbf{GPU}              & NVIDIA Tesla V100 32GB                              \\ \hline
\textbf{RAM}              & 512GB 3200MHz DDR4                                 \\ \hline
\textbf{NIC} & NVIDIA Mellanox ConnectX-5 25Gbps \& ConnectX-6 100Gbps
\\ \hline
\textbf{Software} & Ubuntu 22.04 LTS, NVIDIA Driver 535.183.01, CUDA 12.2\\ \hline
\end{tabular}
}
\end{table}
\end{flushleft}



%% file: prototype.tex
\label{sec:implementation}

\noindent 
\textbf{Constructing Virtual Memory:} As \mytitle{} relies on virtual memory, a large CPU/host memory region is allocated using  ~\textit{malloc} to host application data and registered to RNIC ~\textit{ibv\_reg\_mr} verb with appropriate access flags before the applications start running on GPU. This ensures that the RNIC has the necessary address translations. The access information of host memory which includes remote keys, starting address, and the length of the allocated/allowed region is copied into GPU memory for easy access from GPU threads during runtime.


\noindent 
\textbf{Access to RNIC from GPU Threads:} To achieve high throughput, GPU threads are expected to solve page fault requests fast. The main high-latency part is where the GPU threads start inserting the requests into the send queue and update the doorbell as these steps involve memory accesses. Prior work(s) ~\cite{silberstein2016gpunet} proposed to map the RNIC resources such as QP and CQ buffers and doorbell registers to boost the performance of network-related applications on GPUs. However, GPU threads need to access these resources through PCIe bus, and due to high access latencies and lack of efficient concurrency, this approach cannot achieve high throughput with low granularities (page size in our case).  Instead, we allocate completion and queue pair buffers on GPU memory using ~\textit{cudaMalloc} and associate these buffers with the RNIC resources, namely, ~\textit{ibv\_cq} and ~\textit{ibv\_qp}. To this end, we modify the ~\textit{rdma\_qp\_create} and ~\textit{ibv\_cq\_create} functions in the newest Mellanox Infiniband driver to include user-defined buffers similar to the GPUrdma ~\cite{daoud2016gpurdma}. This makes sure the GPU has access to the RNIC necessary buffers in its memory.

To give GPU access to the doorbell registers, we map them to CUDA address space using  GPUDirect Async ~\cite{nvidia2024magnumio}. For this, we register the doorbells using ~\textit{cudaHostRegister} API with ~\textit{cudaHostRegisterIoMemory} flag to map them to GPU's BAR space and get device pointers using ~\textit{cudaGetDevicePointer}, enabling GPU threads to ring them on demand.


\noindent 
\textbf{Discussion:}
~\textbf{Multi-kernel support}. In the current \mytitle{} implementation, the GPU memory is statistically allocated. The mapping of the pages from host memory to GPU memory happens dynamically during application run time. This paper evaluates \mytitle{} mostly on single-kernel applications and multi-kernel with kernels launched back-to-back. However, the current implementation can be easily extended for concurrently launched kernels in which a stream of concurrent kernels reads from the large dataset on demand.


~\textbf{Multi-GPU processing.} \mytitle{} currently supports 2 GPUs and 2 NICs enabling multi-GPU co-processing in the system. The GPUs can share the NICs for data transfer and concurrently work on the data independently without requiring the programmer to manually create and transfer partitions to the GPUs, separately to amplify the read throughput and access data on demand. 


\input{eval_GDR}

\subsection{Limitations}
We have developed \mytitle{} software prototype on an r7525-type node in Cloudlab ~\cite{Duplyakin+:ATC19}. The system configuration is shown in Figure ~\ref{fig:system_topology}. The performance of the \mytitle{} design is directly related to and limited by components used in this topology. The node consists of 2 GPUs and 2 NICs that are connected to the root complex through dedicated bridges. 
While the kernel is executing on the GPU, any page fault will send/trigger a request to the NIC (1). Upon fetching the page fault work request, NIC sends a request to memory and the data is sent to memory (2). Upon the request, memory DMA sends the requested data to NIC (3). Once NIC receives the data, it finally sends the requested page to GPU memory (3). The first drawback in this transfer mechanism is the interruption of the NIC as it is located on the data path. Another downside is that the data incoming to and outgoing from the RNIC shares the same PCIe bridge channel decreasing the one-directional bandwidth to half of the available bandwidth. To overcome this issue, one possible solution is to modify the requests from NIC to memory such that the page is delivered to GPU memory directly from system memory. However, with modern NICs, this approach is not feasible, as it necessitates modifications to the closed-source NIC firmware. Alternatively, we use both RNICs available on the node for data transfers, increasing the transfer throughput to the maximum available bandwidth as shown in Figure \ref{fig:GDR_Comparison}.



%% file: eval_GDR.tex
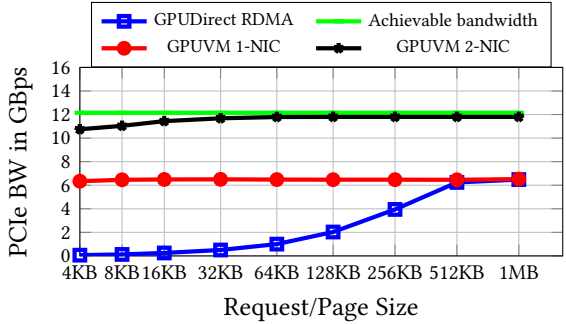
\begin{figure}[t]
    \begin{tikzpicture}
        \begin{axis}[
            xlabel={Request/Page Size},
            ylabel={PCIe BW in GBps},
            grid=major,
            legend style={
                at={(0.5,1.00)},
                anchor=south,
                legend columns=2, 
                /tikz/every even column/.append style={column sep=0.2cm}, 
                font=\scriptsize 
            },
            xmin =0,
            xtick=      {0,   1.5, 3,    5,    7,    9,    11.2,  13.4,  15.6}, 
            xticklabels={4KB, 8KB, 16KB, 32KB, 64KB, 128KB, 256KB, 512KB, 1MB}, 
            tick label style={font=\footnotesize}, 
            width=0.45\textwidth, 
            height=0.23\textwidth, 
            ytick={0, 2, 4, 6, 8, 10, 12, 14, 16},
            ymax=16, 
            ymin=0,
        ]
        \addplot[
            color=blue,
            mark=square,
            ultra thick
        ] coordinates {
            (0, 0.064) (1.5, 0.125) (3, 0.251) (5, 0.503) (7, 1.002) (9, 2.028) (11.2, 3.945) (13.4, 6.237) (15.6, 6.480) 
        };
        \addlegendentry{GPUDirect RDMA}

        \addplot[
            color=green,
            mark=-,
            ultra thick
        ] coordinates {
            (0, 12.15)
            (1.5, 12.15) (3, 12.15) (5, 12.15) (7, 12.15) (9, 12.15) (11.2, 12.15) (13.4, 12.15) (15.6, 12.15) 
        };
        \addlegendentry{Achievable bandwidth}

        \addplot[
            color=red,
            mark=*,
            ultra thick
        ] coordinates {
            (0, 6.34) (1.5, 6.46) (3, 6.49) (5, 6.50) (7, 6.48) (9, 6.47) (11.2, 6.47) (13.4, 6.46) (15.6, 6.52) 
        };
        \addlegendentry{\mytitle{} 1-NIC}

        \addplot[
            color=black,
            mark=star,
            ultra thick
        ] coordinates {

            (0, 10.75) (1.5, 11.04) (3, 11.44) (5, 11.68) (7, 11.79) (9, 11.8) (11.2, 11.8) (13.4, 11.8) (15.6, 11.8) 
        };
        \addlegendentry{\mytitle{} 2-NIC}



        \end{axis}
    \end{tikzpicture}
    \caption{Achieved PCIe bandwidth (BW) with \mytitle{} and GPUDirect RDMA ~\cite{GPUDirectRDMA}. \mytitle{} can achieve the maximum bandwidth even at 4KB page size. GPUDirect RDMA can only saturate the interconnect after 512KB granularity.}
    \label{fig:GDR_Comparison}
    \vspace{-0.25in}
\end{figure}

%% file: evaluation.tex

\input{graph_datasets}

In this section, we provide the performance evaluation of \mytitle{} for different benchmarks and applications and compare the results with other state-of-the-art. The experiments have been conducted on r7225 nodes of Cloudlab ~\cite{Duplyakin+:ATC19}, a testbed with the configuration shown in Table ~\ref{tab:table_hardware}.

\input{eval_csr_uvm_rdma_graph}

\subsection{Comparison with GPUDirect RDMA}
First, we compare the performance of \mytitle{} with NVIDIA GPUDirect RDMA ~\cite{GPUDirectRDMA} on a simple data transfer benchmark with different request sizes. Request size is given as input in scatter-gather entry and defines the length of the data to be transferred from host/system memory to GPU memory through RNIC. The benchmark includes the transfer of 12GB of data from host memory to GPU memory with different request sizes ranging from 4KB to 1MB. In the case of GDR, the transfers are initiated from 16 concurrent threads in the CPU. For \mytitle{}, each GPU warp initiates a transfer with the same request size of GDR corresponding to consecutive addresses. \mytitle{} launches 16 warps in each of the 84 SMs at the same time. Each warp is assigned a page.
The benchmark is conducted using a single NIC and 2 NICs. As shown in Figure ~\ref{fig:GDR_Comparison}, \mytitle{} can achieve the max usable bandwidth available for transfers through a single NIC which is 6.5 GBps even with a 4KB page size. Conversely, GDR can reach the maximum available bandwidth utilization after requests of 512KB. \mytitle{} can keep a stable performance with almost all page sizes and fully utilize the PCIe 3 bandwidth with 2 NICs.

In the case of UVM, we optimize the access memory access patterns such that each warp accesses 64KB of consequent data and uses memory hints (memadvise) for optimization. UVM does not allow flexible page sizes. Therefore, we can report the average throughput only which is 6GBps achieving only 50\% of the available bandwidth.

\begin{figure}[b]
\vspace{-10pt}
    \centering
    \includegraphics[width=6.3cm]{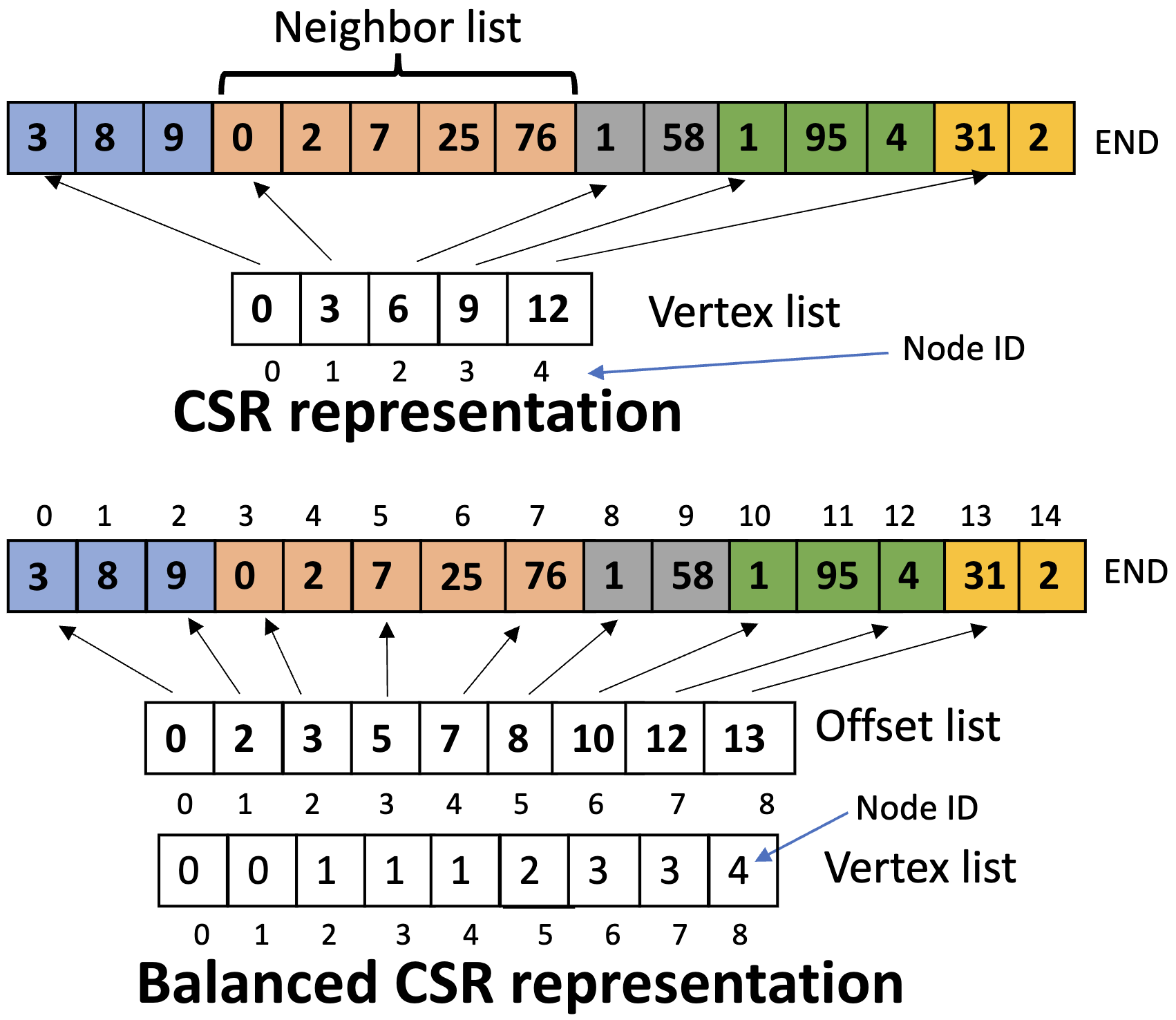}
    \caption{CSR vs Balanced CSR.  Balanced CSR enables multiple threads to concurrently work on the \textbf{same} neighbor list reducing the traversal latency for \mytitle{}.}
    \label{fig:csr_representation}
\end{figure}

\subsection{Graph Workloads}
In this section, we evaluate \mytitle{} on different graph analytics workloads with various graphs listed in Table ~\ref{tab:graph_datasets}. All evaluated graphs are from the SuiteSparse Matrix collection ~\cite{davis2011university}.  The main goal of the \mytitle{} paged-memory system is to provide better results over OS-supported UVM solutions and possibly better results than approaches using bulk transfer. To this end, we choose a well-optimized open-source UVM solution ~\cite{min2020emogi} as a UVM baseline (~\textbf{U}). Additionally, we also compare results with Subway ~\cite{sabet2020subway} which first partitions and preprocesses the graph to create subgraphs of smaller graphs on the CPU and transfers the partitions to the GPU for traversal.

For graph workloads, we choose Breadth First Search (~\textbf{BFS}), Connected Components (~\textbf{CC}), and Single-source-shortest-path (~\textbf{SSSP}). For BFS and SSSP, the application is executed with more than 100 source vertices with at least 2 neighbors and we take the average of the execution times. The reported time for each approach is the average application execution time and UVM memory advise API time if applied. As Section \S\ref{sec:system_overview} describes, the integration of the existing state-of-the-art implementation into \mytitle{} is easy.
We use 1 NIC and 2 NICs with 84 queue-pairs of 64 entries.

\input{table_subway}

For \mytitle{} (~\textbf{G}), we provide 2 versions of implementation; naive and optimized kernels for graph workloads, and compare them against UVM alternatives ~\cite{min2020emogi}
and partitioning-preprocessing-based work ~\cite{sabet2020subway}. The naive approach (\textbf{1N}) involves CSR representation of graphs and uses only 1 NIC. Through experiments with the naive approach, we notice that the imbalance in the number of neighbor lists can become a bottleneck for \mytitle{}. Therefore, for optimized \mytitle{} (\textbf{2N}), we create a new representation called Balanced CSR and use 2 NICs. We use this representation mainly for graphs that have vertices with very large neighbor lists. For example, graph FS has a maximum number of degrees ~5200 and GU has a maximum degree of 68. However, GK and MO have a maximum degree of around $7.5M$ and $2.1M$ neighbors, respectively. This means the number of page faults per some thread(s) will be considerably high, resulting in page fault serialization. To prevent this serialization, we introduce Balanced CSR representation, a modified version of CSR that stores edges in chunks of equal numbers as shown in Figure ~\ref{fig:csr_representation}. This new representation ensures an equal amount of computation and a fairly equal number of page faults per worker. The graph applications get considerable performance improvement with \mytitle{} system when represented in Balaced CSR. The memory overhead that the new representation brings can be ignored as it is up to 400 MB for the graphs in Table ~\ref{tab:graph_datasets}. 

The baseline UVM solution ~\cite{min2020emogi} involves optimizations for coalesced memory accesses and assigning each warp some number of vertices (instead of a single thread single vertex model) to traverse. UVM approach is also optimized (\textbf{wm}) for read-only accesses by setting UVM memory as ~\textit{cudaMemAdviseSetReadMostly} using ~\textit{cudaMemAdvise()}. With this flag, the UVM driver will make read-only copies of the pages on GPU memory and increase the application performance by around 25\% over the one without memadvise (\textbf{nm}). Although the application runtime decreases significantly, setting this memory advice hint adds substantial initial delay which is also reported in our evaluation results and is not included in the speedup calculation. UVM performance does not get any noticeable improvement with Balanced CSR representation as it has speculative prefetching and batching in the GPU fault buffer.

\input{slowdown_QP}

~\textbf{Results.} We provide a pair of timing results for both UVM and \mytitle{}. 
Figure \ref{fig:bfs_cc} shows that \mytitle{}'s performance nearly doubles when using the second NIC, achieving an average 1.4$\times$ improvement for BFS and 1.5$\times$ for CC. 

While \mytitle{} is based on transfers of 8KB pages, the UVM is based on a page size of a minimum of 64KB (4KB fault + 60KB speculative prefetching). Although UVM benefits much from the asynchronous transfer of 60KB, \mytitle{} relies entirely on hardware-mediated transfer and involves small page fault overhead. As a result, \mytitle{} with 2 RNICs can achieve 1.4$\times$ and 1.5$\times$ speedup on BFS and CC graph applications, respectively, over well-optimized UVM solutions.

~\textbf{Comparison with Subway.} As \mytitle{} provides a transparent coding experience for application developers by abstracting away data transfers, it is essential that its performance matches or exceeds that of solutions utilizing explicit data transfers, such as \textit{cudaMemcpy}. In this sense, we compare the results of the BFS and CC applications with Subway ~\cite{sabet2020subway}. Subway provides a solution for large graph traversal by first pinning and processing the graph on the CPU to create smaller subgraphs of active nodes that can fit into GPU memory and transferring these subgraphs to GPU for traversal. As Table ~\ref{table:results} shows \mytitle{} can achieve an average speedup of 1.4$\times$ for BFS and 1.6$\times$ for CC over Subway ~\cite{sabet2020subway}.
Subway is limited to graphs of less than $2^{32}$ vertices, it cannot support the~\textbf{MO} graph.

\input{eval_sssp}

\input{different_apps}

\textbf{Impact of queue count.} To understand the effect of the number of queues on the performance, we measure the slowdown in the performance of CC and BFS as the number of queues is changed. As shown in Figure \ref{fig:queue_number_effect}, the applications achieve optimal performance as the queue count exceeds 48. 

\input{oversubscription_graphs}

\textbf{SSSP with limited GPU memory.} To understand the effect of GPU memory limitation on application performance, we evaluate \mytitle{} and UVM for SSSP with GPU memory limited to 16GB. As the GPU memory is limited, both approaches need to evict the data from the GPU memory to bring a new page. As Figure ~\ref{fig:pagefault} presents \mytitle{} achieves an overall speedup of 1.9$\times$ on SSSP performance on limited GPU memory. As the eviction size in UVM is coarse-grain (2MB), it can evict the data that has not been accessed by GPU threads and might be needed later. However, since \mytitle{} efficiently leverages the reference counter for each page and has an eviction size of 8KB, \mytitle{} reduces the redundant data transfer by 1.8$\times$.

\subsection{Transfer-bound Applications}

In this section, we evaluate performance on CUDA benchmarks: MVT (matrix-vector transpose), ATAX (matrix transpose and vector multiplication), BIGC (big compute), and VA (vectoradd). The performance of these applications depends on efficient data transfer as they involve simple computations. The workloads in this benchmark suite can fit into GPU memory.
MVT, ATAX, and BIGC benchmarks ~\cite{gu2020uvmbench} involve the transpose of a matrix that requires memory accesses through the columns of the matrix reducing the spatial locality. For these applications, \mytitle{} performs around 4$\times$ better than UVM when using two NICs, and around 2$\times$ with one NIC achieving much better PCIe utilization as shown in Figure ~\ref{fig:transfer_bound}. 
Vectoradd given in Listing~\ref{lst:vector_addition} adds two vectors with two billion floating-point elements and stores the results in another vector. \mytitle{} can provide just over 2$\times$ speedup and better PCIe utilization as depicted in Figure ~\ref{fig:transfer_bound}. When we restrict the GPU memory for VA, we obtain around 1.7$\times$ speedup over UVM. Lower speedup result is achieved because \mytitle{} is not optimized for efficient writebacks under GPU memory pressure; specifically, we have not yet implemented asynchronous write-back, resulting in increased latencies on writeback operations.  

\subsection{Oversubscription Analysis}
In this analysis, the goal is to observe the performance slowdown of different applications when the GPU memory gets smaller than the workload size. For this, we artificially limit the GPU memory and keep the workload size fixed. The pressure on the GPU memory compared to the workload size can be defined as in (\ref{equation::1})
\begin{equation}
\frac{\text{Workload Size}}{\text{Available GPU memory}} - 1
\label{equation::1}
\end{equation}

One of the goals of \mytitle{} is to alleviate the programmers from having to handle manual data partitioning and transfer by offering efficient oversubscription of GPU memory with stable performance. The existing approach is UVM oversubscription which is based on an OS page fault handler. Prior works ~\cite{shao2022oversubscribing, allen2024fine, knap2019performance} have shown the degradation of application performance with UVM oversubscription. The performance degradation is directly related to the inefficient eviction mechanism which has a large page size of 2MB and can often evict the pages even before they are accessed by the GPU threads under memory pressure.

\input{rapids}

\mytitle{}, on the other hand, implements a FIFO-based reference priority eviction mechanism as described in Section \S\ref{sec:paged_memory}. This eviction scheme happens when there is not enough memory for \mytitle{} runtime and the thread needs to get a new page number that maps the page in host memory to a page in GPU memory. In this case, the runtime atomically gets a new page number and checks the reference counter. Once the counter becomes zero; no thread needs to access the page,  it is safe to evict the page and the page number is assigned to the new CPU page.

We run several different applications under different oversubscription levels. The key observation from Figure \ref{fig:memorygrams} is that \mytitle{} can provide consistent and predictable performance even under highly pressured GPU memory. However, with UVM, the performance is mostly dependent on the data access patterns in the application. UVM can slow down the performance of graph applications by 4$\times$ compared to around 2$\times$ slowdown of \mytitle{}. With MVT, ATAX, and BIGC, UVM introduces exponential slowdown since these applications access the matrix or array along the columns (no spatial locality) making no use of speculative prefetching. \mytitle{}, on the other hand, can keep the performance stable by introducing up to 2$\times$ slowdown. 

The similar jumps in the slowdown for all applications in \mytitle{} happen because the leader threads call eviction functions that introduce additional latency to page fault time. 

\subsection{Query Evaluation Benchmarks}
For query evaluation, we compare the performance of RAPIDS by NVIDIA and custom query search with UVM and \mytitle{}. RAPIDS is a framework to make query searches over the datasets using GPU. We provide 5 queries for Chicago Taxi Trips Dataset ~\cite{chicago_taxi_trips}. The main question in the query is to find ~\textit{"The average dollar per mile a driver makes in trips that take longer than 9000 seconds".} To find the ultimate result, firstly, (Q1) we find the total miles in such trips (>9000 seconds), secondly, we find the total fares (Q2), then total extras (negatives) (Q3), then the total tips (Q4), and finally, total tolls (Q5). For all comparisons, the dataset is loaded into the system memory. To optimize the run time of RAPIDS, pinned buffers are employed. \mytitle{} and UVM rely on host memory as the backup and make on-demand page requests to the data in host memory.

~\textbf{Results}. Figure ~\ref{fig:rapids} shows that UVM performs on average 1.5$\times$ and 3$\times$ slower than RAPIDS and \mytitle{}, respectively, and gives no advantage for I/O amplification. \mytitle{} benefits from the high-throughput smaller granularity page transfers leading to a performance improvement of over 1.5$\times$ with one NIC and 2.5$\times$ with two NICs over RAPIDS (Q5). Although RAPIDS can benefit from high-bandwidth data transfers due to pinned buffers, it needs to transfer the whole data (entire columns) for processing as it lacks on-demand access. Therefore, RAPIDS cannot improve I/O amplification. UVM cannot deliver any I/O enhancement as the page size is high (64KB). However, \mytitle{} halves I/O amplification (redundant data transfer) for sparsity of 0.08\%.


\input{register_usage}



%% file: graph_datasets.tex
\begin{table}[b!]
    \centering
    \small
    \textbf{$|$E$|$: number of edges, $|$V$|$: number of vertices}
    \resizebox{0.40\textwidth}{!}{
    \begin{tabular}{cccccc}
        \toprule
        \multirow{2}{*}{\textbf{Dataset Name}} & \multirow{2}{*}{\textbf{Abbr}} & \multirow{2}{*}{\textbf{$|$E$|$}} & \multirow{2}{*}{\textbf{$|$V$|$}} & \multicolumn{2}{c}{\textbf{Size (GB)}} \\
        \cmidrule(lr){5-6}
        & & & & \textbf{Edges} & \textbf{Weights} \\
        \midrule
        GAP-Urand ~\cite{kunegis2013konect} & GU & 4.29B & 134.2M & 16.0 & 16.0 \\
        \midrule
        GAP-Kron ~\cite{kunegis2013konect} & GK & 4.23B & 134.2M & 15.7 & 15.7 \\
        \midrule
        Friendster ~\cite{yang2012defining} & FS & 3.61B & 65.6M & 13.5 & 13.5 \\
        \midrule
        MOLIERE ~\cite{sybrandt2017moliere} & MO & 6.67B & 30.2M & 24.8 & 24.8 \\
        \bottomrule
    \end{tabular}
    }
    
    \caption{Description of graph datasets.}
    \label{tab:graph_datasets}
    \vspace{-30pt}
\end{table}


%% file: eval_csr_uvm_rdma_graph.tex
\begin{figure*}[t]
    \pgfmathsetmacro{\b}{0.0}
    \pgfmathsetmacro{\t}{0.2}
    \pgfmathsetmacro{\T}{0.4}

    \begin{tikzpicture}
      \begin{axis}[
        ybar stacked,
        x axis line style={opacity=0},
        ylabel={Time (s)}, 
        tickwidth=0pt,
        bar width=6pt, 
        xtick={\b, \b+\t*1+\T*0, \b+\t*2+\T*0, \b+\t*3+\T*0, 
               \b+\t*3+\T*1, \b+\t*4+\T*1, \b+\t*5+\T*1, \b+\t*6+\T*1,
               \b+\t*6+\T*2, \b+\t*7+\T*2, \b+\t*8+\T*2, \b+\t*9+\T*2,
               \b+\t*9+\T*3, \b+\t*10+\T*3, \b+\t*11+\T*3, \b+\t*12+\T*3,  
               \b+\t*12+\T*4, \b+\t*13+\T*4, \b+\t*14+\T*4, \b+\t*15+\T*4, 
               \b+\t*15+\T*5, \b+\t*16+\T*5, \b+\t*17+\T*5, \b+\t*18+\T*5, 
               \b+\t*18+\T*6, \b+\t*19+\T*6, \b+\t*20+\T*6, \b+\t*21+\T*6,
               \b+\t*21+\T*7, \b+\t*22+\T*7, \b+\t*23+\T*7, \b+\t*24+\T*7},
         xticklabels={~\textbf{U\_FS\_wm}, ~\textbf{U\_FS\_nm}, ~\textbf{G\_FS\_1N}, ~\textbf{G\_FS\_2N},  
                      ~\textbf{U\_GK\_wm}, ~\textbf{U\_GK\_nm}, ~\textbf{G\_GK\_1N}, ~\textbf{G\_GK\_2N}, 
                      ~\textbf{U\_GU\_wm}, ~\textbf{U\_GU\_nm}, ~\textbf{G\_GU\_1N}, ~\textbf{G\_GU\_2N},
                      ~\textbf{U\_MO\_wm}, ~\textbf{U\_MO\_nm}, ~\textbf{G\_MO\_1N}, ~\textbf{G\_MO\_2N},
                     ~\textbf{U\_FS\_wm}, ~\textbf{U\_FS\_nm}, ~\textbf{G\_FS\_1N}, ~\textbf{G\_FS\_2N}, 
                     ~\textbf{U\_GK\_wm}, ~\textbf{U\_GK\_nm}, ~\textbf{G\_GK\_1N}, ~\textbf{G\_GK\_2N}, 
                     ~\textbf{U\_GU\_wm}, ~\textbf{U\_GU\_nm}, ~\textbf{G\_GU\_1N}, ~\textbf{G\_GU\_2N},
                     ~\textbf{U\_MO\_wm}, ~\textbf{U\_MO\_nm}, ~\textbf{G\_MO\_1N}, ~\textbf{G\_MO\_2N}},
        xticklabel style={font=\small, rotate=90, anchor=east}, 
        enlarge x limits=0.2,
        ymin=0,
        ymax=12,
        ytick={0, 2, 4,  6, 8, 10, 12, 14, 16},
        yticklabel={\pgfmathprintnumber{\tick}},
        xticklabel style={text height=1ex},
        legend style={at={(0.5,1.25)}, anchor=north, legend columns=-1, draw=none, fill=none},
        ymajorgrids,
        width=18cm,
        height=3.5cm,
        enlarge x limits={abs=0.1}, 
        clip=false, 
        ]

        \def\xbfsU{\b} 
        \def\xbfsR{\b+\t*2}
        \def\xbfsOptU{0} 
        \def\xbfsOptR{\b+\t*2}

        \def\xccU{\b+\t*12+\T*4} 
        \def\xccR{\b+\t*14+\T*4} 
        \def\xccOptU{\b+\t*12+\T*4} 
        \def\xccOptR{\b+\t*14+\T*4} 
        
        \def\xssspbase{0} 
        \def\xoffset{\t}
        \def\xoffseT{\T}
        
        \addplot coordinates {
         (\xbfsU, 3.1) (\xbfsOptU+\xoffset*1, 3.53)  
         (\xbfsR, 3.656) (\xbfsOptR+\xoffset*1, 2.15)   
         
         (\xbfsU+\xoffset*3+\xoffseT*1, 3.2) (\xbfsOptU+\xoffset*4+\xoffseT*1, 4.2) 
         (\xbfsR+\xoffset*3+\xoffseT*1, 5.358) (\xbfsOptR+\xoffset*4+\xoffseT*1, 2.04) 
         
         (\xbfsU+\xoffset*6+\xoffseT*2, 3.41) (\xbfsOptU+\xoffset*7+\xoffseT*2, 4.49) 
         (\xbfsR+\xoffset*6+\xoffseT*2, 3.736) (\xbfsOptR+\xoffset*7+\xoffseT*2, 2.8) 

         (\xbfsU+\xoffset*9+\xoffseT*3, 4.46) (\xbfsOptU+\xoffset*10+\xoffseT*3, 6.06) 
         (\xbfsR+\xoffset*9+\xoffseT*3, 6.42) (\xbfsOptR+\xoffset*10+\xoffseT*3, 3.75) 
         (\xccU, 3.329)       (\xccOptU+\xoffset*1, 4.096)   
         (\xccR, 4.87) (\xccOptR+\xoffset*1, 2.395)   
         
         (\xccU+\xoffset*3+\xoffseT*1, 4.063)       (\xccOptU+\xoffset*4+\xoffseT*1, 4.97)  
         (\xccR+\xoffset*3+\xoffseT*1, 10.2) (\xccOptR+\xoffset*4+\xoffseT*1, 2.81)  
         
         (\xccU+\xoffset*6+\xoffseT*2, 4.862)    (\xccOptU+\xoffset*7+\xoffseT*2, 5.677) 
         (\xccR+\xoffset*6+\xoffseT*2, 4.26) (\xccOptR+\xoffset*7+\xoffseT*2, 2.81) 

         (\xccU+\xoffset*9+\xoffseT*3, 5.065)    (\xccOptU+\xoffset*10+\xoffseT*3, 7.016) 
         (\xccR+\xoffset*9+\xoffseT*3, 12) (\xccOptR+\xoffset*10+\xoffseT*3, 4.132) 

        };

        \addplot coordinates {

         (\xbfsU, 0) (\xbfsOptU+\xoffset*1, 0)  
         (\xbfsR, 0) (\xbfsOptR+\xoffset*1, 0)   
         
         (\xbfsU+\xoffset*3+\xoffseT*1, 0) (\xbfsOptU+\xoffset*4+\xoffseT*1, 0) 
         (\xbfsR+\xoffset*3+\xoffseT*1, 0) (\xbfsOptR+\xoffset*4+\xoffseT*1, 0) 
         
         (\xbfsU+\xoffset*6+\xoffseT*2, 0) (\xbfsOptU+\xoffset*7+\xoffseT*2, 0) 
         (\xbfsR+\xoffset*6+\xoffseT*2, 0) (\xbfsOptR+\xoffset*7+\xoffseT*2, 0) 

         (\xbfsU+\xoffset*9+\xoffseT*3, 0) (\xbfsOptU+\xoffset*10+\xoffseT*3, 0) 
         (\xbfsR+\xoffset*9+\xoffseT*3, 0) (\xbfsOptR+\xoffset*10+\xoffseT*3, 0) 
         (\xccU, 0)       (\xccOptU+\xoffset*1, 0)   
         (\xccR, 0) (\xccOptR+\xoffset*1, 0)   
         
         (\xccU+\xoffset*3+\xoffseT*1, 0)       (\xccOptU+\xoffset*4+\xoffseT*1, 0)  
         (\xccR+\xoffset*3+\xoffseT*1, 0) (\xccOptR+\xoffset*4+\xoffseT*1, 0)  
         
         (\xccU+\xoffset*6+\xoffseT*2, 0)    (\xccOptU+\xoffset*7+\xoffseT*2, 0) 
         (\xccR+\xoffset*6+\xoffseT*2, 0) (\xccOptR+\xoffset*7+\xoffseT*2, 0) 

         (\xccU+\xoffset*9+\xoffseT*3, 0)    (\xccOptU+\xoffset*10+\xoffseT*3, 0) 
         (\xccR+\xoffset*9+\xoffseT*3, 0) (\xccOptR+\xoffset*10+\xoffseT*3, 0) 
        };
        
        \addplot coordinates {
         (\xbfsU, 0) (\xbfsOptU+\xoffset*1, 0)  
         (\xbfsR, 0) (\xbfsOptR+\xoffset*1, 0)   
         
         (\xbfsU+\xoffset*3+\xoffseT*1, 0) (\xbfsOptU+\xoffset*4+\xoffseT*1, 0) 
         (\xbfsR+\xoffset*3+\xoffseT*1, 0) (\xbfsOptR+\xoffset*4+\xoffseT*1, 0) 
         
         (\xbfsU+\xoffset*6+\xoffseT*2, 0) (\xbfsOptU+\xoffset*7+\xoffseT*2, 0) 
         (\xbfsR+\xoffset*6+\xoffseT*2, 0) (\xbfsOptR+\xoffset*7+\xoffseT*2, 0) 

         (\xbfsU+\xoffset*9+\xoffseT*3, 0) (\xbfsOptU+\xoffset*10+\xoffseT*3, 0) 
         (\xbfsR+\xoffset*9+\xoffseT*3, 0) (\xbfsOptR+\xoffset*10+\xoffseT*3, 0) 
         (\xccU, 0)       (\xccOptU+\xoffset*1, 0)   
         (\xccR, 0) (\xccOptR+\xoffset*1, 0)   
         
         (\xccU+\xoffset*3+\xoffseT*1, 0)       (\xccOptU+\xoffset*4+\xoffseT*1, 0)  
         (\xccR+\xoffset*3+\xoffseT*1, 0) (\xccOptR+\xoffset*4+\xoffseT*1, 0)  
         
         (\xccU+\xoffset*6+\xoffseT*2, 0)    (\xccOptU+\xoffset*7+\xoffseT*2, 0) 
         (\xccR+\xoffset*6+\xoffseT*2, 0) (\xccOptR+\xoffset*7+\xoffseT*2, 0) 

         (\xccU+\xoffset*9+\xoffseT*3, 0)    (\xccOptU+\xoffset*10+\xoffseT*3, 0) 
         (\xccR+\xoffset*9+\xoffseT*3, 0) (\xccOptR+\xoffset*10+\xoffseT*3, 0) 
        };

        \addplot coordinates {
        
         (\xbfsU, 2.25) (\xbfsOptU+\xoffset*1, 0)  
         (\xbfsR, 0) (\xbfsOptR+\xoffset*1, 0)   
         
         (\xbfsU+\xoffset*3+\xoffseT*1, 2.65) (\xbfsOptU+\xoffset*4+\xoffseT*1, 0) 
         (\xbfsR+\xoffset*3+\xoffseT*1, 0) (\xbfsOptR+\xoffset*4+\xoffseT*1, 0) 
         
         (\xbfsU+\xoffset*6+\xoffseT*2, 2.7) (\xbfsOptU+\xoffset*7+\xoffseT*2, 0) 
         (\xbfsR+\xoffset*6+\xoffseT*2, 0) (\xbfsOptR+\xoffset*7+\xoffseT*2, 0) 

         (\xbfsU+\xoffset*9+\xoffseT*3, 4.24) (\xbfsOptU+\xoffset*10+\xoffseT*3, 0) 
         (\xbfsR+\xoffset*9+\xoffseT*3, 0) (\xbfsOptR+\xoffset*10+\xoffseT*3, 0) 
         (\xccU, 2.25)       (\xccOptU+\xoffset*1, 0)   
         (\xccR, 0) (\xccOptR+\xoffset*1, 0)   
         
         (\xccU+\xoffset*3+\xoffseT*1, 2.65)       (\xccOptU+\xoffset*4+\xoffseT*1, 0)  
         (\xccR+\xoffset*3+\xoffseT*1, 0) (\xccOptR+\xoffset*4+\xoffseT*1, 0)  
         
         (\xccU+\xoffset*6+\xoffseT*2, 2.7)    (\xccOptU+\xoffset*7+\xoffseT*2, 0) 
         (\xccR+\xoffset*6+\xoffseT*2, 0) (\xccOptR+\xoffset*7+\xoffseT*2, 0) 

         (\xccU+\xoffset*9+\xoffseT*3, 4.24)    (\xccOptU+\xoffset*10+\xoffseT*3, 0) 
         (\xccR+\xoffset*9+\xoffseT*3, 0) (\xccOptR+\xoffset*10+\xoffseT*3, 0) 
        };

        \legend{Application Runtime,, , Memadvise}
      \end{axis}

      \def\Mul{0.38}
      \draw[dashed] ({\b+\t*12+\T*12+1.0}, -1.8) -- ({\b+\t*12+\T*12+1.0}, 2.0);  
      
      \def\xoffsetbfs{0.4} 
      \def\yoffsetbfs{-1.8}
      \node[below=0cm, font=\small] at (\xoffsetbfs+3.6,\yoffsetbfs+0.0) {~\textbf{BFS}};

      \def\xoffsetcc{3.0} 

      \def\xoffsetcc{5.6} 
      \node[below=0cm, font=\small] at (\xoffsetcc+7.0,\yoffsetbfs+0.0) {~\textbf{CC}};

      \def\xoffsetcc{8.2} 


    \end{tikzpicture}
    
    \caption{Graph workload evaluation results. '\textbf{U}': UVM, '\textbf{wm}': with memory advise hint (applies to only UVM), '\textbf{nm}': no memory advise hint, '\textbf{G}': \mytitle{}, '\textbf{1N}': 1 NIC in CSR, '\textbf{2N}': 2 NIC in Balanced CSR. \mytitle{} implementation uses 8KB-sized pages. \mytitle{}'s overall performance is 1.4$~\times$ for BFS and 1.5$~\times$ for CC better than UVM solution.}
    \label{fig:bfs_cc}
\end{figure*}

%% file: table_subway.tex
\begin{table}[t]
\centering
\small
\resizebox{0.38\textwidth}{!}{
\begin{tabular}{ccccc}
\toprule
\textbf{Benchmark} & \textbf{Graph} & \multicolumn{2}{c}{\textbf{Total Time (s)}} & \textbf{Speedup} \\ 
                   &                & \textbf{Subway}    & \textbf{\mytitle{}}    &                  \\ \midrule
\multirow{3}{*}{BFS}  & GK & 3.86s  & 2.04 & 1.89$\times$ \\ 
                      & GU & 3.13s  & 2.80 & 1.12$\times$ \\
                      & FS & 2.52  & 2.15 & 1.17$\times$ \\ \midrule
\multirow{3}{*}{CC}   & GK & 4.73s  & 2.81 & 1.68$\times$ \\ 
                      & GU & 5.21s  & 2.80 & 1.86$\times$ \\
                      & FS &  3.24s  & 2.40 & 1.35$\times$ \\ \bottomrule
\end{tabular}
}
\caption{Performance comparison to Subway~\cite{sabet2020subway}}
\label{table:results}
\vspace{-25pt}
\end{table}

%% file: slowdown_QP.tex
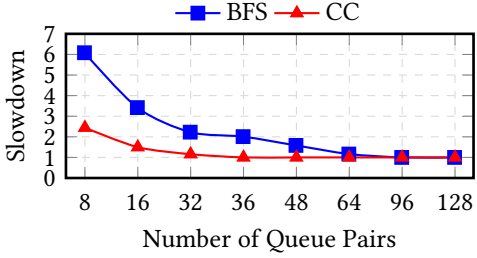
\begin{figure}[t]
    \pgfmathsetmacro{\b}{0.0}  
    \pgfmathsetmacro{\t}{0.01}  
    \pgfmathsetmacro{\T}{1.2}  

    \begin{tikzpicture}
      \begin{axis}[
        xlabel={Number of Queue Pairs},
        ylabel={Slowdown},
        xtick={\b, \b+\T*1, \b+\T*2, \b+\T*3, \b+\T*4, \b+\T*5, \b+\T*6, \b+\T*7},
        xticklabels={8, 16, 32, 36, 48, 64, 96, 128},
        xticklabel style={yshift=-0.5ex}, 
        enlarge x limits=0.05, 
        ymin=0,
        ymax=7,
        ytick={0, 1, 2, 3, 4, 5, 6, 7},
        yticklabel={\pgfmathprintnumber{\tick}},
        legend style={at={(0.5,1.3)}, anchor=north, legend columns=-1, draw=none, fill=none},
        grid=major,
        major grid style={dashed, gray!30}, 
        width=7.0cm,
        height=3.5cm,
        clip=false,
        mark options={scale=1.2},
        cycle list name=color list, 
        line width=1pt, 
        smooth, 
      ]

        \addplot[mark=square*, thick, color=blue] coordinates {
         (\b - \t*1, 18675/3070)   
         (\b - \t*1 + \T*1, 10496/3070) 
         (\b - \t*1 + \T*2, 6837/3070) 
         (\b - \t*1 + \T*3, 6170/3070) 
         (\b - \t*1 + \T*4, 4861/3070) 
         (\b - \t*1 + \T*5, 3588/3080) 
         (\b - \t*1 + \T*6, 1) 
         (\b - \t*1 + \T*7, 3070/3070) 
        };

        \addplot[mark=triangle*, thick, color=red] coordinates {
         (\b + \t*0, 8481/3471)
         (\b + \t*0 + \T*1, 5196/3471)
         (\b + \t*0 + \T*2, 4015/3471)
         (\b + \t*0 + \T*3, 1)
         (\b + \t*0 + \T*4, 1)
         (\b + \t*0 + \T*5, 1)
         (\b + \t*0 + \T*6, 1)
         (\b + \t*0 + \T*7, 3471/3471)
        };

        \legend{BFS, CC}
      \end{axis}
    \end{tikzpicture}
    
    \caption{Sensitivity to number of QPs and CQs}
    \label{fig:queue_number_effect}
    \vspace{-15pt}
\end{figure}

    


%% file: eval_sssp.tex
\begin{figure}[t]
    \pgfmathsetmacro{\b}{0.0}
    \pgfmathsetmacro{\t}{0.2}
    \pgfmathsetmacro{\T}{0.4}

    \begin{tikzpicture}
      \begin{axis}[
        ybar,
        x axis line style={opacity=0},
        ylabel={Speedup}, 
        tickwidth=0pt,
        bar width=6pt, 
        xtick={\b, \b+\t*3+\T*1, \b+\t*6+\T*2, \b+\t*9+\T*3}, 
        xticklabels={FS, GK, GU, MO}, 
        xticklabel style={yshift=-0.5ex}, 
        enlarge x limits={0.1, abs=0.2}, 
        ymin=0,
        ymax=3, 
        ytick={0, 0.2, 0.4, 0.6, 0.8, 1, 1.2, 1.4, 1.6, 1.8, 2, 2.2, 2.4, 2.6, 2.8, 3, 3.2, 3.4, 3.6, 3.8, 4.0}, 
        yticklabels={0, , , , , 1, , , , , 2, , , , , 3, , , , , 4}, 
        ymajorgrids=true,
        grid style={black!10}, 
        ymajorgrids=true,
        grid style={black!5, ultra thin},
        extra y ticks={0, 1, 2, 3}, 
        extra y tick style={grid style={black!100, thin}},
        xticklabel style={text height=1ex, rotate=0}, 
        legend style={at={(0.25,1.08)}, anchor=south, legend columns=2, draw=none, fill=none}, 
        width=7.0cm,
        height=3.5cm,
        clip=false, 
        enlarge x limits={abs=0.4},
        ]

        \addplot [
          fill=blue!70, 
          draw=black!100 
        ] coordinates {
         (\b, 37995.57/37995.57) 
         (\b+\t*3+\T*1, 56326.04/56326.04) 
         (\b+\t*6+\T*2, 191730.22/191730.22) 
         (\b+\t*9+\T*3, 29837.01/29837.01) 
        };
        \addlegendentry{UVM}; 

        \addplot [
          fill=brown!80, 
          draw=black!100 
        ] coordinates {
         (\b+\t*0, 37995.57/16286.13)  
         (\b+\t*3+\T*1, 56326.04/31832.457031) 
         (\b+\t*6+\T*2, 191730.22/136459.906250) 
         (\b+\t*9+\T*3, 29837.01/14757.89) 
        };
        \addlegendentry{\mytitle{}}; 

      \end{axis}

      \begin{axis}[
        axis y line*=right, 
        axis x line=none, 
        ylabel={I/O Amp. Ratio}, 
        ymin=0,
        ymax=3, 
        ytick={0, 1, 2, 3}, 
        xtick=\empty, 
        width=7.0cm,
        height=3.5cm,
        enlarge x limits={abs=0.4}, 
        clip=false,
        legend style={at={(0.75,1.08)}, anchor=south, legend columns=1, draw=none, fill=none}, 
        ]

        \addplot[green!80, thick, mark=triangle, mark options={scale=1.0}] coordinates {
         (\b, 1.9)  
         (\b+\t*3+\T*1, 2.62)  
         (\b+\t*6+\T*2, 1.41)  
         (\b+\t*9+\T*3, 1.38)  
        };
        \addlegendentry{I/O Amp. Ratio}; 
      \end{axis}
    \end{tikzpicture}

    \caption{SSSP with 16GB GPU memory.}
    \label{fig:pagefault}
\end{figure}
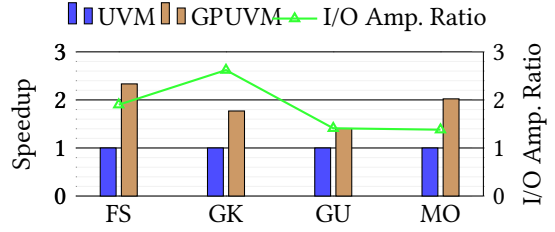

%% file: different_apps.tex
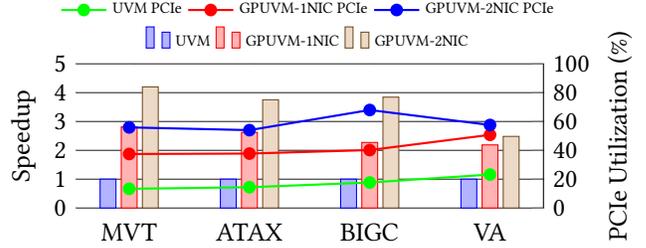
\begin{figure}[t]
    \pgfmathsetmacro{\b}{0.0}  
    \pgfmathsetmacro{\t}{0.1}  
    \pgfmathsetmacro{\T}{0.9}  

    \begin{tikzpicture}
      \begin{axis}[
        ybar, 
        x axis line style={opacity=0}, 
        ylabel={Speedup},
        tickwidth=0pt,
        bar width=6pt,
        xtick={\b, \b+\T*1, \b+\T*2, \b+\T*3, \b+\T*4},
        xticklabels={MVT, ATAX, BIGC, VA, app5},
        xticklabel style={yshift=-0.5ex}, 
        enlarge x limits=0.05, 
        ymin=0,
        ymax=5,
        ytick={0, 1, 2, 3, 4, 5},
        yticklabel={\pgfmathprintnumber{\tick}},
        legend style={at={(0.5,1.3)}, anchor=north, legend columns=-1, draw=none, fill=none},
        grid style={black!90}, 
        ymajorgrids=true,
        grid style={black!90, ultra thin},
        width=7.8cm, 
        height=3.5cm,
        clip=false,
        enlarge x limits={abs=0.40},
      ]

        \addplot coordinates {
         (\b - \t*0, 10017/10017)   
         (\b - \t*0 + \T*1, 9265/9265) 
         (\b - \t*0 + \T*2, 7543/7543) 
         (\b - \t*0 + \T*3, 5751/5751) 
        };

        \addplot coordinates {
         (\b + \t*0, 10017/3560) 
         (\b + \t*0 + \T*1, 9265/3532) 
         (\b + \t*0 + \T*2, 7543/3319) 
         (\b + \t*0 + \T*3, 5751/2628) 
        };

        \addplot coordinates {
         (\b + \t*0, 10017/2385) 
         (\b + \t*0 + \T*1, 9265/2469) 
         (\b + \t*0 + \T*2, 7543/1962) 
         (\b + \t*0 + \T*3, 5751/2316) 
        };

        \legend{\tiny UVM, \tiny \mytitle{}-1NIC, \tiny \mytitle{}-2NIC}
      \end{axis}
      
      \begin{axis}[
        axis y line*=right, 
        ylabel={ PCIe Utilization (\%) }, 
        ymin=0, ymax=100, 
        ytick={0, 20, 40, 60, 80, 100}, 
        yticklabel={\pgfmathprintnumber{\tick}}, 
        legend style={at={(0.5,1.5)}, anchor=north, legend columns=-1, draw=none, fill=none},
        width=7.8cm, 
        height=3.5cm, 
        hide x axis, 
        axis x line=none, 
        xshift=0.0cm, 
        enlarge x limits={abs=0.40},
      ]

      \addplot[green, thick, mark=*] coordinates {
        (\b - \t*0, {(16/10.017)*100/12})   
         (\b - \t*0 + \T*1, {(16/9.265)*100/12}) 
         (\b - \t*0 + \T*2, {(16/7.543)*100/12}) 
         (\b - \t*0 + \T*3, {(16/5.751)*100/12}) 
      };

      \addplot[red, thick, mark=*] coordinates {
        (\b + \t*0, {(16/3.560)*100/12}) 
         (\b + \t*0 + \T*1, {(16/3.532)*100/12}) 
         (\b + \t*0 + \T*2, {(16/3.319)*100/12}) 
         (\b + \t*0 + \T*3, {(16/2.628)*100/12}) 
      };

      \addplot[blue, thick, mark=*] coordinates {
        (\b + \t*0, {(16/2.385)*100/12}) 
         (\b + \t*0 + \T*1, {(16/2.469)*100/12}) 
         (\b + \t*0 + \T*2, {(16/1.962)*100/12}) 
         (\b + \t*0 + \T*3, {(16/2.316)*100/12}) 
      };

      \legend{ \tiny UVM PCIe, \tiny \mytitle{}-1NIC PCIe, \tiny  \mytitle{}-2NIC PCIe}
      
      \end{axis}
    \end{tikzpicture}
    
    \caption{Performance (bars) and PCIe Utilization (lines) for UVM and \mytitle{}}
    \label{fig:transfer_bound}
    \vspace{-10pt}
\end{figure}

%% file: oversubscription_graphs.tex
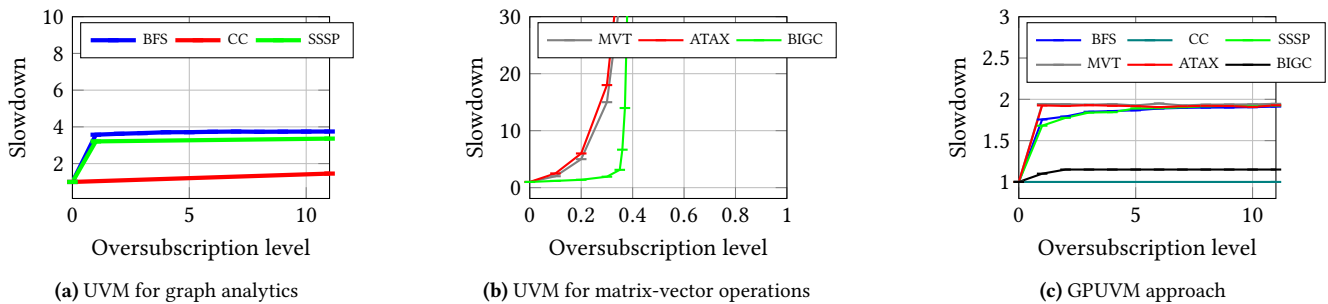
\begin{figure*}[h]
    \centering 
    
    \subfloat[UVM for graph analytics]{
        \small
        \begin{tikzpicture}
            \begin{axis}[
                xlabel={Oversubscription level},
                ylabel={Slowdown},
                grid=major,
                legend pos=north west,
                legend columns=-1,
                xmin =0,
                xmax = 11,
                width=5cm,
                height=4cm,
                ymax=10
            ]
            \addplot[
                color=blue,
                mark=-,
                ultra thick
            ] coordinates {
                (0, 3130/3130) (1,11182/3130) (2,11342/3130) (3,11461/3130) (4,11605/3130) (5, 11616/3130) (6, 11683/3130) (7, 11723/3130) (8,11679/3130) (9, 11704/3130) (10,11715/3130) (11,11729/3130)
            };
            \addlegendentry{\tiny BFS}
            \addplot[color=red, mark=-, ultra thick] coordinates {
                (0,3331/3331) 
 (11, 4849/3331)
            };
            \addlegendentry{\tiny CC}
            \addplot[color=green, mark=-, ultra thick] coordinates {
                (0,3441/3441) 
                (1, 11043/3441) 
 (11,  11564/3441)
            };
            \addlegendentry{\tiny SSSP}
            \end{axis}
        \end{tikzpicture}
    }
    \hfill
    \subfloat[UVM for matrix-vector operations]{
        \small
        \begin{tikzpicture}
            \begin{axis}[
                xlabel={Oversubscription level},
                ylabel={Slowdown},
                grid=major,
                legend pos=north west, 
                legend columns=-1,
                xmin =0,
                xmax = 1,
                width=5cm,
                height=4cm,
                ymax=30
            ]
            \addplot[color=gray, mark=-, thick] coordinates {
                (0, 1/1) (0.1, 2/1) (0.2, 5/1) (0.3, 15/1) (0.4, 50/1)
            };
            \addlegendentry{\tiny MVT}
            \addplot[color=red, mark=-, thick] coordinates {
                (0, 1/1) (0.1, 2.5/1) (0.2, 6/1) (0.3, 18/1) (0.4, 60/1)
            };
            \addlegendentry{\tiny ATAX}

            \addplot[color=green, mark=-, thick] coordinates {
                (0, 1/1) (0.1, 8938/7545) (0.2, 10407/7545) (0.3, 14497/7545) (0.35, 23593/7545) (0.36, 50267/7545) (0.37, 105433/7545) (0.4, 529923/7545)
            };
            \addlegendentry{\tiny BIGC}
            \end{axis}
        \end{tikzpicture}
    }
    \hfill
    \subfloat[\mytitle{} approach]{
        \small
        \begin{tikzpicture}
            \begin{axis}[
                xlabel={Oversubscription level},
                ylabel={Slowdown},
                grid=major,
                legend pos=north west,
                legend columns=3,
                xmin =0,
                xmax = 11,
                width=5cm,
                height=4cm,
                ymax=3
            ]
            \addplot[color=blue, mark=-, thick] coordinates {
                (0, 7590/7590) (1,13329/7590) (2,13600/7590) (3,14037/7590) (4,14101/7590) (5,14173/7590) (6,14334/7590) (7,14393/7590) (8,14420/7590) (9,14430/7590) (10,14459/7590) (11,14512/7590)
            };
            \addlegendentry{\tiny BFS}
            \addplot[color=teal, mark=-, thick] coordinates {
                (0, 9665/9665) (11, 9665/9665)
            };
            \addlegendentry{\tiny CC}
            \addplot[color=green, mark=-, thick] coordinates {
                (0,6092/6093) (1,10241/6093) (2,10820/6093) (3,11211/6093) (4,11239/6093) (5,11513/6093) (6,11541/6093) (7,11604/6093) (8,11673/6093) (9,11704/6093) (10,11708/6093) (11,11730/6093)
            };
            \addlegendentry{\tiny SSSP}
            \addplot[color=gray, mark=-, thick] coordinates { 
            (0, 6656/6656) (1, 12933/6656) (2, 12923/6656) (3, 12876/6656) (4, 12932/6656) (5, 12819/6656) (6, 12978/6656) (7, 12818/6656) (8, 12884/6656) (9, 12894/6656) (10, 12908/6656) (11,  12947/6656)
        };
        \addlegendentry{\tiny MVT}

        \addplot[color=red, mark=-, thick] coordinates { 
            (0, 6646/6646) (1, 12874/6683) (2, 12828/6683) (3, 12896/6683) (4, 12852/6683) (5, 12815/6683) (6, 12734/6683) (7, 12801/6683) (8, 12839/6683) (9, 12825/6683) (10, 12747/6683) (11,  12886/6683)
        };
        \addlegendentry{\tiny ATAX}

        \addplot[color=black, mark=-, thick] coordinates { 
            (0, 1.0/1) (1, 1.1/1) (2, 1.15/1) (3, 1.15/1) (4, 1.15/1) (5, 1.15/1) (6, 1.15/1) (7, 1.15/1) (8, 1.15/1) (9, 1.15/1) (10, 1.15/1) (11,  1.15/1)
        };
        \addlegendentry{\tiny BIGC}
        
            \end{axis}
        \end{tikzpicture}
    }

    \caption{Effect of oversubscription on GPU memory for different algorithms with UVM and \mytitle{}.}
    \label{fig:memorygrams}
\end{figure*}

%% file: rapids.tex
\definecolor{emerald}{RGB}{80,200,120} 
\definecolor{midnightblue}{RGB}{25,25,112} 
\definecolor{skyblue}{rgb}{0.529, 0.808, 0.922}

\begin{figure}[t]
    \pgfmathsetmacro{\b}{0.0}  
    \pgfmathsetmacro{\t}{0.1}  
    \pgfmathsetmacro{\T}{0.9}  

    \begin{tikzpicture}
      \begin{axis}[
    ybar, 
    x axis line style={opacity=0}, 
    ylabel={Speedup},
    tickwidth=0pt,
    bar width=4pt,
    xtick={\b, \b+\T*1, \b+\T*2, \b+\T*3, \b+\T*4},
    xticklabels={Q1, Q2, Q3, Q4, Q5},
    xticklabel style={, yshift=-0ex}, 
    enlarge x limits=0.05, 
    ymin=0,
    ymax=5,
    ytick={0, 1, 2, 3, 4, 5}, 
    yticklabel={\pgfmathprintnumber{\tick}}, 
    extra y tick labels={}, 
    legend style={at={(0.5,1.5)}, anchor=north, legend columns=-1, draw=none, fill=none},
        ymajorgrids=true,
    width=7.9cm, 
    height=3.7cm,
    clip=false,
    enlarge x limits={abs=0.4},
    legend image code/.code={%
        \draw[#1] (0cm,-0.1cm) rectangle (0.1cm,0.1cm); 
    },
]

        \addplot[fill=brown!80!white] coordinates {
         (\b + \t*0, 0.369/0.369) 
         (\b + \t*0 + \T*1, 0.516/0.516) 
         (\b + \t*0 + \T*2, 0.668/0.668)
         (\b + \t*0 + \T*3, 0.820/0.820)
         (\b + \t*0 + \T*4, 0.965/0.965)
        };
        \addlegendentry{\tiny UVM}
        
        \addplot [fill=emerald!90!white] coordinates {
         (\b - \t*0, 0.369/0.17)   
         (\b - \t*0 + \T*1, 0.516/0.25) 
         (\b - \t*0 + \T*2, 0.668/0.39) %
         (\b - \t*0 + \T*3, 0.820/0.53) %
         (\b - \t*0 + \T*4, 0.965/0.67) %
        };
        \addlegendentry{\tiny RAPIDS}

        \addplot[fill=midnightblue!60!white] coordinates {
         (\b + \t*0, 0.369/0.177) 
         (\b + \t*0 + \T*1, 0.516/0.204) 
         (\b + \t*0 + \T*2, 0.668/0.287)
         (\b + \t*0 + \T*3, 0.820/0.345)
         (\b + \t*0 + \T*4, 0.965/0.398)
        };
        \addlegendentry{\tiny \mytitle{}-1NIC}
    
        \addplot[fill=skyblue!90!white] coordinates {
         (\b + \t*0, 0.369/0.118) 
         (\b + \t*0 + \T*1, 0.516/0.146) 
         (\b + \t*0 + \T*2, 0.668/0.177) 
         (\b + \t*0 + \T*3, 0.820/0.220) 
         (\b + \t*0 + \T*4, 0.965/0.241) 
        };
        \addlegendentry{\tiny \mytitle{}-2NIC}


      \end{axis}

      \begin{axis}[
    axis y line*=right, 
    axis x line=none, 
    ymin=0,
    ymax=10, 
    ylabel={I/O Amp. Factor},
    width=7.9cm,
    height=3.7cm,
    clip=false,
    ytick={0, 2, 4, 6, 8, 10}, 
    yticklabel={\pgfmathprintnumber{\tick}}, 
    legend style={at={(0.5,1.3)}, anchor=north, legend columns=-1, draw=none, fill=none},
]

        \addplot[
          color=red,
          mark=square*,
          mark size=1.5pt,
          line width=0.5pt,
        ] coordinates {
         (\b - \t*0, 1.44 )   
         (\b - \t*0 + \T*1, 1.89 ) 
         (\b - \t*0 + \T*2, 2.33 ) %
         (\b - \t*0 + \T*3, 2.78 ) %
         (\b - \t*0 + \T*4, 3.22 ) %
        };
        \addlegendentry{\tiny \mytitle{} Amp. Factor}
        
        \addplot[
          color=green,
          mark=square*,
          mark size=1.5pt,
          line width=0.5pt,
        ] coordinates {
         (\b - \t*0, 2 )   
         (\b - \t*0 + \T*1, 3 ) 
         (\b - \t*0 + \T*2, 4 ) %
         (\b - \t*0 + \T*3, 5 ) %
         (\b - \t*0 + \T*4, 6 ) %
        };

        \addlegendentry{\tiny Rapids Amp. Factor}

        \addplot[
          color=black,
          mark=triangle*,
          mark size=1.5pt,
          line width=0.5pt,
        ] coordinates {
         (\b - \t*0,  2.0-0.15)   
         (\b - \t*0 + \T*1,  3.0-0.15) 
         (\b - \t*0 + \T*2,  4.0-0.15) %
         (\b - \t*0 + \T*3,  4.82-0.15) %
         (\b - \t*0 + \T*4, 5.9 -0.15) %
        };
        \addlegendentry{\tiny UVM Amp. Factor}
        
      \end{axis}
    \end{tikzpicture}
    
    \caption{Comparison of RAPIDS, UVM, and \mytitle{} for query evaluations. \mytitle{} uses 4KB pages. UVM uses a minimum page size of 64KB. ( 0.08\% sparsity)}
    \label{fig:rapids}
    \vspace{-20pt}
\end{figure}
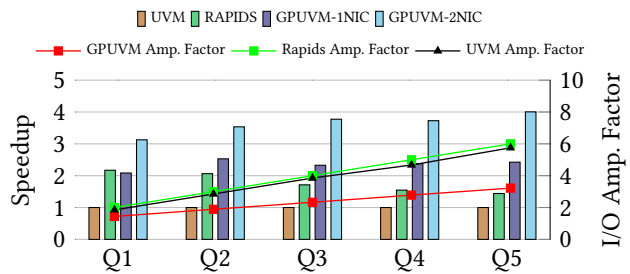

%% file: register_usage.tex
\begin{figure}[t]
    \pgfmathsetmacro{\b}{0.0}
    \pgfmathsetmacro{\t}{0.2}
    \pgfmathsetmacro{\T}{0.4}

    \begin{tikzpicture}
      \begin{axis}[
        ybar,
        x axis line style={opacity=0},
        ylabel={Register Total}, 
        tickwidth=0pt,
        bar width=6pt, 
        xtick={\b, \b+\t*3+\T*1, \b+\t*6+\T*2, \b+\t*9+\T*3, \b+\t*12+\T*4}, 
        xticklabels={BFS, CC, RAPIDS(Q5), MVT, ATAX},
        enlarge x limits={0.1, abs=0.2}, 
        ymin=0,
        ymax=200, 
        ytick={0, 20, 40, 60, 80, 100, 120, 140, 160, 180, 200, 220, 240, 260, 280, 300}, 
        yticklabels={0, , , , , 100, , , , , 200, , , , , 300}, 
        ymajorgrids=true,
        grid style={black!10}, 
        ymajorgrids=true,
        grid style={black!10, ultra thin},
        extra y ticks={0, 100, 200, 300}, 
        extra y tick style={grid style={black!50, thin}},
        xticklabel style={font=\footnotesize, text height=1ex, rotate=0, xshift=4pt}, 
        legend style={at={(0.5,1.08)}, anchor=south, legend columns=2, draw=none, fill=none}, 
        width=8.0cm,
        height=2.6cm,
        clip=false, 
        enlarge x limits={abs=0.4},
        nodes near coords, 
        every node near coord/.append style={font=\footnotesize} 
        ]

        \addplot [
          fill=blue!70, 
          draw=black!100 
        ] coordinates {
         (\b, 34) 
         (\b+\t*3+\T*1, 96) 
         (\b+\t*6+\T*2, 28) 
         (\b+\t*9+\T*3, 24) 
         (\b+\t*12+\T*4, 20) 
        };
        \addlegendentry{UVM}; 

        \addplot [
          fill=brown!80, 
          draw=black!100 
        ] coordinates {
         (\b+\t*1, 48)  
         (\b+\t*4+\T*1, 114) 
         (\b+\t*7+\T*2, 32) 
         (\b+\t*10+\T*3, 72) 
         (\b+\t*13+\T*4, 72) 
        };
        \addlegendentry{\mytitle{}}; 

      \end{axis}
      
    \end{tikzpicture}

    \caption{Register use per thread for UVM and \mytitle{}. No register spilling occurs for any application.}
    \label{fig:register_usage}
    \vspace{-0.2in}
\end{figure}
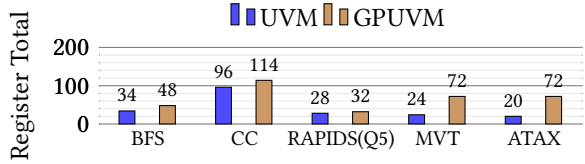

%% file: discussion.tex
\mytitle{} allows GPU threads to directly interface with the NIC.  Specifically, both page mappings and page tables are stored in GPU memory, enabling faster access for GPU threads. In total, \mytitle{} incurs memory overhead of up to 8MB for page tables and mappings, and 2MB for NIC control resources. In addition, as Figure ~\ref{fig:register_usage} illustrates using \mytitle{} on GPUs does not cause any register spilling for the applications studied in this paper. 
Furthermore, the current implementation of \mytitle{} supports a system of up to two GPUs and two NICs. The multi-NICs can facilitate fast parallel data communication between the accelerators and system memory, allowing for efficient multi-GPU collaboration for applications that require big datasets and can be run on multiple GPUs in a parallel fashion, making \mytitle{} a scalable solution.

%% file: related_work.tex

\textbf{Unified memory and storage}. Prior works ~\cite{ziabari2016umh, ganguly2019interplay, yu2020coordinated, kim2020batch, koukos2016building, choi2022memory, go2023early, li2019compiler, markthub2018dragon, zhang2023g10, park2023uvmmu} have been proposed to enable efficient memory access through unified memory and storage with UVM. 
For example, Choi et al. ~\cite{choi2022memory} proposes a new system for efficient unified memory systems for multi-GPU systems. Koukos et al. ~\cite{koukos2016building} propose a novel scheme to facilitate heterogeneous systems (CPU-GPU) with unified virtual memory. Ziabari et al. ~\cite{ziabari2016umh} proposes a new hardware-based unified memory hierarchy for multi-GPU systems. Zhang et al. ~\cite{zhang2023g10} presents a unified GPU memory and storage architecture for tensor migration in deep learning workloads. Markthub et al. ~\cite{markthub2018dragon} extends the UVM to be able to make page-fault requests to storage devices. Other works ~\cite{kim2020batch, yu2020coordinated, go2023early} mainly leverage software modifications to improve the UVM performance. The common part among these works is that they rely on the host OS page fault handler for data migration. Therefore, performance is limited due to the lack of page fault handling parallelization and OS involvement delay. In addition, the recently introduced Grace-Hopper Superchip ~\cite{fusco2024understanding, schieffer2024harnessing} supports unified memory with hardware page tables and creates coherent CPU-GPU memory by bringing together the Grace CPU and Hopper GPU through NVLink-C2C~\cite{nvidia_grace_hopper}. However, it relies on a new and different CPU design that supports NVLink connection becoming an expensive hardware-dependant solution. 

~\textbf{Enabling direct storage access for larger memory}. Some prior works ~\cite{shahar2016activepointers, qureshi2023gpu, silberstein2013gpufs, vesely2018generic} enable direct storage access for GPUs. Silberstein et al. ~\cite{silberstein2013gpufs} proposed POSIX-like file system APIs for GPU programs by integrating CPU's cache buffer into GPU memory. Active pointers by Shahar et al. ~\cite{shahar2016activepointers} are abstractions similar to memory-map to enable GPU threads to directly access the storage devices. Qureshi et al. ~\cite{qureshi2023gpu} recently proposed a software cache in GPU memory to enable on-demand, fine-grain, and high-throughput access to the storage. \mytitle{} differs from these works as it delivers efficient on-demand access to system memory, rather than storage and provides efficient oversubscription of GPU memories.  

\mytitle{} can be thought of as a form of memory disaggregation, but driven by GPUs. Prior works~\cite{dragojevic2014farm, ruan2020aifm, shan2018legoos} aim to provide efficient fine-grain manipulation of remote data. GPUVM is similar to these works as all of these works rely on user space to manage the memory accesses. However, GPUVM is different from these works as it focuses on efficient memory management for GPUs within a single node. 

Other prior works ~\cite{daoud2016gpurdma, silberstein2016gpunet, hamidouche2020gpu} proposed to enable GPU direct access to the network. GPUrdma ~\cite{daoud2016gpurdma} is a network library for GPUs, while GPUnet ~\cite{silberstein2016gpunet} presents a native GPU networking layer with socket abstraction and high-level networking APIs. We acknowledge all the contributions made by these works. 

\vspace{-0.1in}

%% file: conclusion.tex
In this paper, we address the data transfer bottleneck between the GPU and backup memory (typically located on the CPU), which constrains performance, especially for workloads with datasets that exceed the capacity of GPU memory. We propose a novel approach for allowing the GPU to manage this data transfer directly through RDMA.  Since the CPU motherboard does not support establishing RDMA connections directly, we leverage a Network Interface Card (NIC) to facilitate this direct communication. With this support, we show that GPUVM can substantially outperform UVM especially when the application memory footprint exceeds the available physical memory on the GPU.  


%% file: ref.tex
\bibliography{ref.bib}